\documentclass[preprint,12pt]{elsarticle}

\usepackage{graphicx} % Required for inserting images
\usepackage{amsmath}  % for writing normal text in an equation
\usepackage{amssymb}  % for symbol \triangleq, \mathbb
\usepackage[T1]{fontenc}
\usepackage{tikz}
\usetikzlibrary{shapes.geometric, arrows.meta, positioning, fit, shadows, decorations.pathreplacing}
\usepackage{hyperref}
\usepackage{adjustbox}
\usepackage{placeins} % for figure placement control
\usepackage{booktabs} % for better tables
\usepackage{xcolor}  % for text color
\usepackage[group-separator={,}]{siunitx}  % for separator

\journal{Computers and Chemical Engineering}

\begin{document}

% \input{commands.tex}
% utility
\newcommand{\inb}[1]{\left(#1\right)}
\newcommand{\bm}[1]{\mathbf{#1}}
\newcommand{\colored}[2]{\textcolor{#1}{#2}}
\newcommand{\emptyline}{\vspace{12pt}}
\newcommand{\safespace}[1]{#1~}

\newcommand{\vectorized}[1]{#1}  % {\bm{#1}}

\newcommand{\propertyphase}[2]{#1^{#2}}
\newcommand{\propertyindex}[2]{#1_{#2}}
\newcommand{\propertyphaseindex}[3]{\propertyindex{\propertyphase{#1}{#2}}{#3}}

\newcommand{\best}{*}
\newcommand{\mynum}{N}
\newcommand{\column}{Co}
\newcommand{\component}{C}
\newcommand{\spl}{s}
\newcommand{\refluxratio}{RR}
\newcommand{\boilupratio}{BR}
\newcommand{\Acetone}{Ac}
\newcommand{\azeotrope}{Az}

% already exists -> \begin{quote}...\end{quote} or something like that
% \newcommand{\mycitation}[1]{"\textit{#1}"}  % citation

\newcommand{\ones}{\vectorized{1}}

% \eqref already exists
\newcommand{\figureref}[1]{Fig.~\ref{#1}}
\newcommand{\sectionref}[1]{Sec.~\ref{#1}}
\newcommand{\tabref}[1]{Tab.~\ref{#1}}
\newcommand{\appendixref}[1]{Appendix~\ref{#1}}

% operators
%     use!
% \newcommand{\true}[1]{#1^*}
\newcommand{\estimator}[1]{\hat{#1}}

% states
\newcommand{\vapor}{V}
\newcommand{\liquid}{L}
\newcommand{\critical}{C}
\newcommand{\saturated}{\mathcal{S}}
\newcommand{\saturatedvapor}{\saturated \vapor}
\newcommand{\vaporization}{\Delta \vapor}

% Abbreviations
\newcommand{\modelfluid}{MF}
\newcommand{\margules}{M}
\newcommand{\simplifiedantoine}{SA}
\newcommand{\extendedantoine}{EA}
\newcommand{\entrainer}{E}
\newcommand{\process}{P}

\newcommand{\rigorous}{Rig}
\newcommand{\surrogatemodel}{SM}

\newcommand{\total}{tot}
\newcommand{\reboiler}{Reb}
\newcommand{\condenser}{Con}
\newcommand{\product}{Pro}
\newcommand{\hypothetical}{Hypo}
\newcommand{\candidate}{Can}
\newcommand{\recycle}{Rec}
\newcommand{\feed}{F}
\newcommand{\bottom}{B}
\newcommand{\distillate}{D}
\newcommand{\flowsheet}{F}
\newcommand{\abovefeed}{AF}
\newcommand{\belowfeed}{BF}
\newcommand{\stages}{S}

% variables
\newcommand{\liquidmolarfraction}{\ell}
\newcommand{\liquidmolarfractions}{\vectorized{\liquidmolarfraction}}
\newcommand{\vapormolarfraction}{v}
\newcommand{\vapormolarfractions}{\vectorized{\vapormolarfraction}}
\newcommand{\vaporflow}{V}
\newcommand{\liquidflow}{L}
\newcommand{\temperature}{T}
\newcommand{\pressure}{p}
\newcommand{\molarenthalpy}{h}
\newcommand{\feature}{x}
\newcommand{\target}{y}
\newcommand{\features}{\vectorized{\feature}}
\newcommand{\featurespace}{\mathcal{X}}
\newcommand{\targetspace}{\mathcal{Y}}
\newcommand{\parameter}{\theta}
\newcommand{\activitycoefficient}{\gamma}
\newcommand{\heatduty}{\dot{Q}}
\newcommand{\reboilerduty}{\heatduty^{\reboiler}}
\newcommand{\totalreboilerduty}{\heatduty^{\reboiler, \total}}
\newcommand{\condenserduty}{\heatduty^{\condenser}}
\newcommand{\confidenceinterval}{\text{CI}}

\newcommand{\model}{f}
\newcommand{\loss}{\mathcal{L}}
\newcommand{\constraint}{\model}
\newcommand{\jacobian}{J}

%     maybe use?
% \newcommand{\latentvariable}{z}
% \newcommand{\dataset}{\mathcal{D}}

\newcommand{\nrtlparameters}{\parameter^{\text{NRTL}}}
\newcommand{\extendedantoineparameters}{\parameter^{\text{\extendedantoine}}}
\newcommand{\modelfluidfeature}{\feature^{\text{\modelfluid}}}
\newcommand{\modelfluidfeatures}{\vectorized{\modelfluidfeature}}
\newcommand{\modelfluidparameter}{\parameter^{\modelfluid}}
\newcommand{\modelfluidparameters}{\vectorized{\modelfluidparameter}}

\newcommand{\liquidmolarenthalpy}{\propertyphase{\molarenthalpy}{\liquid}}
\newcommand{\vapormolarenthalpy}{\propertyphase{\molarenthalpy}{\vapor}}
\newcommand{\vaporizationenthalpy}{\propertyphase{\molarenthalpy}{\vaporization}}

% are those actually used?
\newcommand{\saturatedvaportemperature}{\propertyphase{\temperature}{\saturatedvapor}}
\newcommand{\saturatedvaportemperatureof}[1]{\propertyindex{\saturatedvaportemperature}{#1}}
\newcommand{\liquidmolarfractionof}[1]{\propertyindex{\liquidmolarfraction}{#1}}
\newcommand{\vapormolarfractionof}[1]{\propertyindex{\vapormolarfraction}{#1}}
\newcommand{\vaporizationenthalpyof}[1]{\propertyindex{\vaporizationenthalpy}{#1}}

% special short cuts for the appendix
\newcommand{\derivativefeatureof}[1]{\frac{\partial \vapormolarfractionof{#1}}{\partial \liquidmolarfractionof{#1}}}
\newcommand{\derivativefeatureatinfdilution}[2]{\derivativefeatureof{#1}\vert_{#2}}
\newcommand{\activitycoefficientatinfinitedilutionin}[2]{\activitycoefficient_{#1}\vert_{#2}}
\newcommand{\saturatedvaporpressurevariantaofat}[2]{\pressure_{#1, \vapor}^{\saturated, \rigorous}\inb{#2, \extendedantoineparameters}}
\newcommand{\activitycoefficientvariantaofat}[1]{\activitycoefficient_{#1}^{\rigorous}\inb{\liquidmolarfractionof{#1}=0, \temperature\inb{\liquidmolarfractionof{#1}=0}}}

\newcommand{\barunit}{\text{bar}}
\newcommand{\molunit}{\text{mol}}
\newcommand{\molefracunit}{\text{mol}\:\text{mol}^{-1}}
\newcommand{\wattunit}{\text{W}}

\newcommand{\aspen}{Aspen Plus V10}

\tikzstyle{stagenumber} = [font=\bfseries] % define the stagenumber style

\newcommand{\normalstage}[2]{
    \begin{scope}[shift={(0,#2)}]  % apply shift
        \draw (0, 0) rectangle (2, 0.5);
        \draw [->] (0.5, 0.) -- (0.5, -0.5);
        \draw [->] (1.5, 0.5) -- (1.5, 1.);
        \node[stagenumber] at (0.5, 0.25) {$#1$};
        \node[stagenumber] at (1.5, 0.25) {$\temperature^{({#1})}$};
        \node[stagenumber] at (3.5, 0.75) {$\vaporflow^{(\stages^{({#1})})}$, $\vapormolarfractions^{(\stages^{({#1})})}$};
        \node[stagenumber] at (-1.5, -0.25) {$\liquidflow^{(\stages^{({#1})})}$, $\liquidmolarfractions^{(\stages^{({#1})})}$};
    \end{scope}
}

\newcommand{\bottomstage}[2]{
    \begin{scope}[shift={(0,#2)}]  % apply shift
        % draw dot
        \draw (0.5, 0.44) circle (0.05);

        % draw stream arrows
        \draw [->] (0.5, 0.40) -- (0.5, -0.75);
        \draw [->] (0.54, 0.44) -- (1.5, 0.44) -- (1.5, 1.0);
        
        % draw heat exchanger
        \draw (1.0, 0.25) circle (0.35);
        \draw [-] (0.75, -0.35) -- (0.75, 0.35);
        \draw [-] (1.25, -0.35) -- (1.25, 0.35);
        \draw [-] (0.75, 0.35) -- (1., 0.15);
        \draw [-] (1., 0.15) -- (1.25, 0.35);

        % draw description
        \node[stagenumber] at (1.1, -0.6) {$\reboilerduty$};
        \node[stagenumber] at (0.1, 0.3) {$\boilupratio$};
        \node[stagenumber] at (3, 0.75) {$\vaporflow^{({#1})}$, $\vapormolarfraction^{({#1})}$, $\temperature^{({#1})}$};
        \node[stagenumber] at (-1, -0.75) {$\liquidflow^{\bottom}$, $\liquidmolarfraction^{\bottom}$, $\temperature^{\bottom}$};
    \end{scope}
}

\newcommand{\distillatestage}[2]{
    \begin{scope}[shift={(0,#2)}]  % apply shift
        % draw dot
        \draw (1.5, 0.05) circle (0.05);

        % draw stream arrows
        \draw [->] (1.5, 0.1) -- (1.5, 0.8);
        \draw [->] (1.45, 0.05) -- (0.5, 0.05) -- (0.5, -0.5);

        % draw heat exchanger
        \draw (1.0, 0.25) circle (0.35);
        \draw [-] (0.75, 0.15) -- (0.75, 0.85);
        \draw [-] (1.25, 0.15) -- (1.25, 0.85);
        \draw [-] (0.75, 0.15) -- (1., 0.35);
        \draw [-] (1., 0.35) -- (1.25, 0.15);

        % draw description
        \node[stagenumber] at (1.0, 1.2) {$\condenserduty$};
        \node[stagenumber] at (1.9, 0.2) {$\refluxratio$};

        \node[stagenumber] at (-1.4, -0.25) {$\liquidflow^{({#1})}$, $\liquidmolarfraction^{({#1})}$, $\temperature^{({#1})}$};
        \node[stagenumber] at (2.6, 1) {$\liquidflow^{\distillate}$, $\liquidmolarfraction^{\distillate}$, $\temperature^{\distillate}$};
    \end{scope}
}

\newcommand{\feedstage}[2]{
    \begin{scope}[shift={(0,#2)}]  % apply shift
        \draw (0, 0) rectangle (2, 0.5);
        \draw [->] (0.5, 0.) -- (0.5, -0.5);
        \draw [->] (1.5, 0.5) -- (1.5, 1.);
        \draw [->] (-1., 0.25) -- (0., 0.25);
        \node[stagenumber] at (0.5, 0.25) {#1};
        \node[stagenumber] at (1.5, 0.25) {$\temperature^{({#1})}$};
        \node[stagenumber] at (3.5, 0.75) {$\vaporflow^{({#1})}$, $\vapormolarfraction^{({#1})}$};
        \node[stagenumber] at (-1, -0.25) {$\liquidflow^{({#1})}$, $\liquidmolarfraction^{({#1})}$};
        \node[stagenumber] at (-2.5, 0.25) {$\liquidflow_{\feed}$, $\liquidmolarfraction_{\feed}$, $\temperature_{\feed}$};
    \end{scope}
}

\begin{frontmatter}
    \title{Reusable Surrogate Models for Distillation Columns}
    \author[1]{Martin Bubel\corref{cor1}}
    \author[1]{Tobias Seidel}
    \author[1]{Michael Bortz}
    \cortext[cor1]{Corresponding author. Email: martin.bubel@itwm.fraunhofer.de}
    \affiliation[1]{
        organization={Department of Optimization, Fraunhofer Institute for Industrial Mathematics},
        addressline={Fraunhofer-Platz 1},
        city={Kaiserslautern},
        postcode={D-67663},
        state={Rheinland-Pfalz},
        country={Germany}
    }

    \begin{abstract}
        Surrogate modeling is a powerful methodology in chemical process engineering, frequently employed to accelerate optimization tasks.
        Despite their popularity, most surrogate models are trained for a narrow range of fixed chemical systems and operating conditions, which limits their reusability.
        This work introduces a paradigm shift towards reusable surrogates by developing a single model for distillation columns that generalizes across a vast design space.
        The key enabler is a novel ML-fueled modelfluid representation which allows for the generation of datasets of more than $\num{1000000}$ samples.
        This allows the surrogate to generalize not only over column specifications but also over the entire chemical space of homogeneous ternary vapor-liquid mixtures.
        We validate the model's accuracy and demonstrate its practical utility in a case study on entrainer distillation, where it successfully screens and ranks candidate entrainers, significantly reducing the computational effort compared to rigorous optimization.
    \end{abstract}

    % Graphical abstract
    % \begin{graphicalabstract}
    %     \begin{figure}[h]
    %         \input{small-overview.tex}
    %     \end{figure}
    % \end{graphicalabstract}

    % % Research highlights
    % \begin{highlights}
    %     \item Introduces a reusable surrogate model for distillation columns that generalizes across a wide range of column specifications and ternary vapor-liquid mixtures.
    %     \item Leverages a novel modelfluid representation and ML-based property prediction to generate a large, diverse training dataset exceeding one million samples.
    %     \item Demonstrates that a single surrogate model can accurately predict distillation outcomes for any homogeneous ternary mixture, enabling broad applicability.
    %     \item Validates the surrogate model’s performance with rigorous simulation data and a real-world test set, showing high accuracy and robust generalization.
    %     \item Applies the surrogate model to an entrainer selection case study, significantly reducing computational effort in process optimization and screening.
    % \end{highlights}

    % Keywords
\end{frontmatter}

\section{List of Operators, Variables, and Abbreviations}
\noindent
\begin{tabular}{ll}
    \textbf{Operator} & \textbf{Meaning} \\
    $\hat{a}$ & Estimate of $a$ \\
    $a^*$ & Optimal value of $a$ according to some objective function \\
    $a_{i}$ & Index $i$ of $a$ \\
    $a^{\text{phase}}$ & Property $a$ in given phase \\
    $a^{\text{type}}$ & Property $a$ of given type \\
    $\Delta a$ & Difference in $a$ (e.g., error) \\
\end{tabular}

\emptyline
\noindent
\begin{tabular}{ll}
    \textbf{Variable} & \textbf{Meaning} \\
    $\liquidmolarfractionof{i}$ & Molar fraction of component $i$ in the liquid phase \\
    $\vapormolarfractionof{i}$ & Molar fraction of component $i$ in the vapor phase \\
    $\liquidmolarfractions$, $\vapormolarfractions$ & Vector of molar fractions in liquid/vapor phase \\
    $\temperature$ & Temperature \\
    $\pressure$ & Pressure \\
    $\activitycoefficient_i$ & Activity coefficient of component $i$ \\
    $\activitycoefficient_i\vert_j$ & Activity coefficient of component $i$ at infinite dilution in component $j$ \\
    $\refluxratio$ & Reflux ratio \\
    $\spl$ & Bottoms-to-feed split ratio \\
    $\mynum_{\stages}$ & Number of equilibrium stages \\
    $\mynum_{\stages}^{\abovefeed}$, $\mynum_{\stages}^{\belowfeed}$ & Stages above/below feed \\
    $\reboilerduty$ & Reboiler heat duty \\
    $\condenserduty$ & Condenser heat duty \\
    $\vaporizationenthalpyof{i}$ & Enthalpy of vaporization of component $i$ \\
    $\molarenthalpy$ & Molar enthalpy \\
    $\heatduty$ & General heat duty $\dot{Q}$ \\
    $\features$, $\featurespace$ & Feature vector, feature space \\
    $\target$, $\targetspace$ & Target/output, target space \\
    $\parameter$ & Model parameter \\
    $\modelfluidfeatures$ & Modelfluid feature vector \\
    $\modelfluidparameters$ & Modelfluid parameter vector \\
    $\model$ & Surrogate model function \\
    $\loss$ & Loss function $\mathcal{L}$ \\
    $\jacobian$ & Jacobian matrix $J$ \\
    $\confidenceinterval$ & Confidence interval (CI) \\
    $\liquidmolarfractions_{\bottom,i}$ & Bottom stream molar fractions \\
    $\liquidmolarfractions_{\distillate,i}$ & Distillate stream molar fractions \\
\end{tabular}

All variables are in SI units unless stated otherwise.

\emptyline
\noindent

\begin{tabular}{ll}
    \textbf{Abbreviation} & \textbf{Meaning} \\
    ANN & Artificial Neural Network \\
    GNN & Graph Neural Network \\
    ML & Machine Learning \\
    MSE & Mean Squared Error \\
    RMSE & Root Mean Squared Error \\
    MINLP & Mixed-Integer Nonlinear Programming \\
    CAPEX & Capital Expenditure \\
    OPEX & Operational Expenditure \\
    PC-SAFT & Perturbed Chain - Statistical Associating Fluid Theory \\
    MESH & Mass, Equilibrium, Summation, Heat \\
    VLE & Vapor-Liquid Equilibrium \\
    NRTL & Non-Random Two-Liquid (model) \\
    DIPPR & Design Institute for Physical Properties (database) \\
    SMILES & Simplified Molecular Input Line Entry System \\
    CI & Confidence Interval \\
    AI & Artificial Intelligence \\
    \extendedantoine & Extended Antoine (model) \\
    \simplifiedantoine & Simplified Antoine (model) \\
    \modelfluid & Modelfluid \\
    \entrainer & Entrainer \\
    \rigorous & Rigorous (model) \\
    \surrogatemodel & Surrogate Model \\
    \Acetone & Acetone \\
    \azeotrope & Azeotrope \\
    \reboiler & Reboiler \\
    \condenser & Condenser \\
    \bottom & Bottom \\
    \distillate & Distillate \\
    \feed & Feed \\
    \abovefeed & Above Feed \\
    \belowfeed & Below Feed \\
    \stages & Stage \\
\end{tabular}

\section{Introduction}
\label{sec:introduction}
High-fidelity simulation is a cornerstone of modern chemical process engineering, enabling the detailed analysis of complex phenomena, from molecular-level interactions to full-scale plant operations \cite{Forrester2008}.
Distillation columns, in particular, represent a class of unit operations that are critical to the chemical industry but their simulation is known to be computationally expensive.
Rigorous models often involve solving large systems of nonlinear MESH (Mass, Equilibrium, Summation, Heat) equations \cite{Biegler1997}.
Solving these systems usually comes with a high computational effort, which limits the use of high-fidelity models in tasks that require many repeated evaluations, such as large-scale optimization, comprehensive design studies, and rapid exploration of what-if scenarios.

To bridge this gap, surrogate models have become a popular and powerful methodology in the field \cite{McBride2019}.
A surrogate model acts as an efficient, optimization-friendly prediction model of a physical system or its high-fidelity simulation.
It is trained on data from experiments or simulations to mimic the system's behavior, providing near-instantaneous predictions.
Much of the recent research has focused on developing sophisticated surrogate models for specific processes.
For instance, proofs-of-concept have demonstrated their use for prereformer reactors \cite{Schmidt2020}.
Subsequent work extended the applicability to the interconnection of unit operations within a steam methane reforming flowsheet, showing good interpolation capabilities and error dampening \cite{Schack2021, Lueg2021}.
Other applications include using surrogates to accelerate convergence of rigorous simulators \cite{Bubel2021}, enabling interactive process exploration \cite{Baldan2022}, and facilitating adaptive, Bayesian-based optimization of process feasibility and performance \cite{Hoeller2023}.

% this part here has been moved up from ref [#1]
A crucial aspect of surrogate models is their formulation as an \textbf{explicit map}, $\model : \featurespace \to \targetspace$, where the model output $\target \in \targetspace$ is computed in a single forward pass given the model inputs $\feature \in \featurespace$.
This stands in stark contrast to rigorous models that require solving an implicit system of equations.
The benefits of this explicit formulation for optimization are as follows:
Firstly, the computational cost is reduced.
Secondly, analytical derivatives with respect to the input features are readily and cheaply available from modern machine learning frameworks, enabling the use of highly efficient Newton-type optimizers.
Lastly, these surrogates do not suffer from the convergence failures that can affect rigorous simulators during numerical optimization.

However, the application of surrogate models in chemical engineering faces a fundamental limitation: they are typically developed as highly specialized prototypes with a narrow range of validity.
A surrogate trained for a specific distillation column separating a particular chemical mixture quickly becomes obsolete if the column geometry, operating conditions, or, most importantly, the chemical system itself changes.
This system-specific nature has been a major barrier to the transition of surrogate models from academic prototypes to widely applicable, reusable tools for industrial practice.

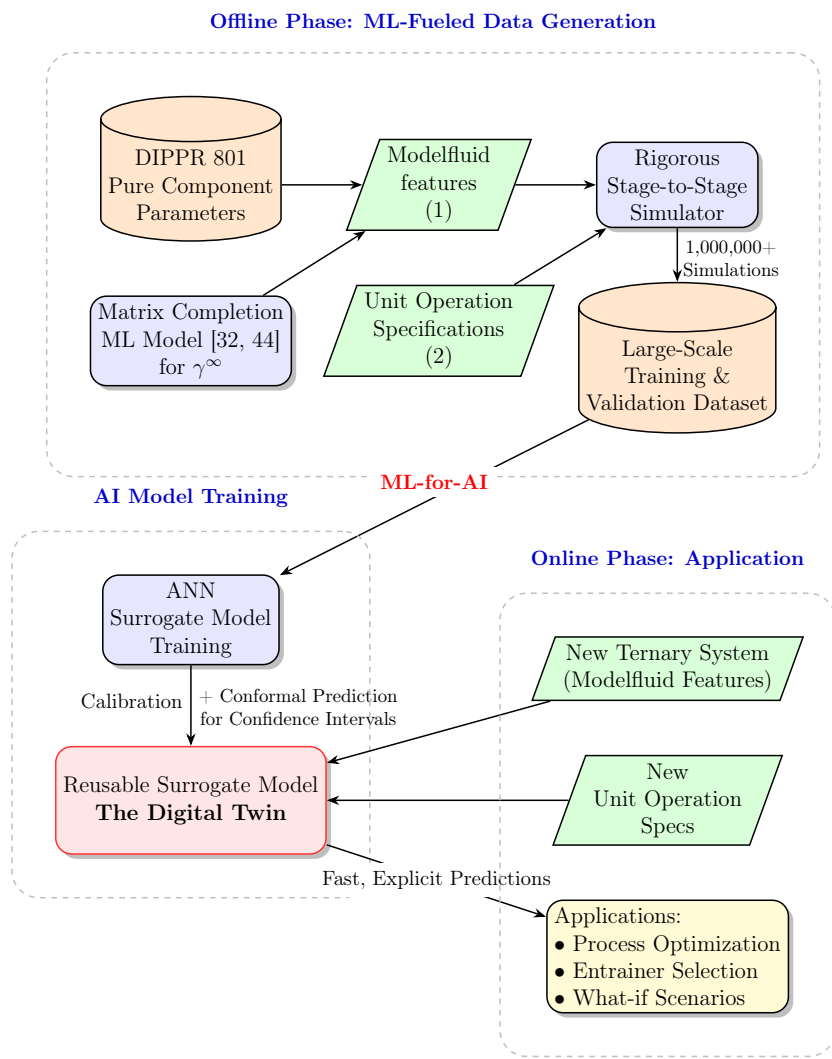
\begin{figure*}[t!]
    \centering
% Use adjustbox to ensure the TikZ picture fits within the text width
\begin{adjustbox}{width=\textwidth,center,scale=0.8}
\begin{tikzpicture}[
    node distance=10mm and 15mm,
    % STYLES
    % Style for main process blocks
    process/.style={
        rectangle, 
        rounded corners=3mm, 
        draw, 
        thick, 
        fill=blue!10, 
        align=center,
        minimum height=15mm, 
        minimum width=30mm,
        drop shadow
    },
    % Style for input/output data
    data/.style={
        trapezium, 
        trapezium left angle=70, 
        trapezium right angle=110,
        draw, 
        thick, 
        fill=green!15, 
        align=center, 
        minimum height=10mm
    },
    % Style for database shapes
    database/.style={
        cylinder, 
        shape border rotate=90, 
        aspect=0.25, 
        draw, 
        thick, 
        fill=orange!20,
        align=center, 
        minimum height=15mm, 
        minimum width=25mm
    },
    % Style for the final "Digital Twin" model
    digitaltwin/.style={
        rectangle, 
        rounded corners=3mm, 
        draw=red!80, 
        thick, 
        fill=red!10, 
        align=center,
        minimum height=20mm, 
        minimum width=40mm,
        drop shadow
    },
    % Style for arrows - uses the 'Stealth' tip from the 'arrows.meta' library
    arrow/.style={
            -Stealth,
            thick
        },
    % Style for descriptive labels
    label/.style={
        text=blue!80!black, 
        font=\small\bfseries
    },
    % Style for background fitting boxes
    fitbox/.style={
        rectangle,
        rounded corners=5mm,
        draw=gray!50,
        thick,
        dashed
    }
]

% == PHASE 1: ML-Fueled Data Generation ==
% Column 1: Inputs
\node (db_dippr) [database] {DIPPR 801 \\ Pure Component \\ Parameters};
\node (ml_mc) [process, below=of db_dippr] {Matrix Completion \\ ML Model \cite{Jirasek2020,Damay2021} \\ for $\gamma^\infty$};

% Column 2: Intermediate Representation
\node (modelfluid) [data, right=of db_dippr] {Modelfluid \\ features \\ \eqref{eq:modelfluid:features}};
\node (unitops) [data, below=of modelfluid] {Unit Operation \\ Specifications \\ \eqref{eq:columnfeatures}};

% Column 3: Simulation
\node (simulator) [process, right=of modelfluid] {Rigorous \\ Stage-to-Stage \\ Simulator};
\node (dataset) [database, below=of simulator] {Large-Scale \\ Training \& \\ Validation Dataset};
\node[align=center, font=\footnotesize, below=of simulator, yshift=9mm, xshift=10mm] {1,000,000+ \\ Simulations};
% Arrows for Phase 1
\draw[arrow] (db_dippr) -- (modelfluid);
\draw[arrow] (ml_mc) -- (modelfluid);
\draw[arrow] (modelfluid) -- (simulator);
\draw[arrow] (unitops) -- (simulator);
\draw[arrow] (simulator) -- (dataset);

% Fit a box around Phase 1
\node (phase1_box) [fitbox, fit=(db_dippr) (ml_mc) (simulator) (dataset), inner sep=8mm] {};
\node[label, above=3mm of phase1_box.north] {Offline Phase: ML-Fueled Data Generation};

% == PHASE 2: AI Model Training ==
% Column 4: Training
\node (training) [process, below=of ml_mc, yshift=-25mm] {ANN \\ Surrogate Model \\ Training};

% Arrow with the "ML-for-AI" story
\draw[arrow] (dataset) -- (training) node[midway, above, label, fill=white, text=red] {ML-for-AI};

% == THE FINAL PRODUCT: The Digital Twin ==
% Column 5: The Digital Twin
\node (dt_model) [digitaltwin, below=of training, yshift=-5mm] {Reusable Surrogate Model \\ \textbf{The Digital Twin}};

% Arrow to the final model
\draw[arrow] (training) -- node[midway, above, font=\small, xshift=-11mm, yshift=-2mm] {Calibration} (dt_model);
% Add the conformal prediction note
\node[align=center, font=\footnotesize, below=of training.south, xshift=20mm, yshift=7mm] {+ Conformal Prediction \\ for Confidence Intervals};

% Fit a box around Phase 2
\node (phase2_box) [fitbox, fit=(training) (dt_model), inner sep=8mm] {};
\node[label, above=3mm of phase2_box.north] {AI Model Training};

% == PHASE 3: Application / Usage ==
% This part shows how the final model is used
\node (user_input_ops) [data, right=of dt_model, xshift=30mm] {New \\ Unit Operation \\ Specs};
\node (user_input_fluid) [data, above=of user_input_ops] {New Ternary System \\ (Modelfluid Features)};
\node (applications) [process, below=of user_input_ops, fill=yellow!20, align=left] {Applications: \\ \textbullet{} Process Optimization \\ \textbullet{} Entrainer Selection \\ \textbullet{} What-if Scenarios};

% % Arrows for Phase 3
\draw[arrow] (dt_model) -- (applications) node[midway, above, font=\small, fill=white, yshift=-3mm] {Fast, Explicit Predictions};
\draw[arrow] (user_input_fluid) -- (dt_model);
\draw[arrow] (user_input_ops) -- (dt_model);

% % Fit a box around Phase 3
\node (phase3_box) [fitbox, fit=(user_input_fluid) (applications), inner sep=8mm] {};
\node[label, above=3mm of phase3_box.north] {Online Phase: Application};

\end{tikzpicture}
\end{adjustbox}
\caption{Schematic overview of the proposed workflow. A machine learning model (Matrix Completion) is used to generate a large dataset (the "ML" part), which then enables the training of a generalizable Artificial Neural Network surrogate (the "AI" part). The final, calibrated model serves as a reusable Digital Twin for rapid process optimization and design space exploration.}
    \label{fig:workflow_schematic}
\end{figure*}

This work introduces a shift towards \textbf{reusable surrogate models} for distillation columns, fundamentally addressing the challenge of limited generalizability.
We demonstrate that a single surrogate model can be trained to have a vast range of validity, not only across different column specifications---such as flow rates, reflux ratio, pressure, number of stages, and feed location---but also, most significantly, across the entire chemical space of homogeneous vapor-liquid ternary mixtures.
To achieve this, a single surrogate model is trained so that it can predict the process outcome of a distillation column over a wide range of specifications for any homogeneous ternary vapor-liquid mixture.
The key enabler for this is the use of a \textbf{novel modelfluid representation}, which we introduced in a previous publication \cite{Bubel2025c}.
This representation, combined with a machine learning-fueled dataset generation strategy, allows the surrogate model to learn the underlying relationships between fluid properties and column behavior, rather than simply memorizing the behavior of one specific system.
A schematic overview of this entire workflow is presented in \figureref{fig:workflow_schematic}.
% ref [#1]

In this paper, we develop and validate this new class of generalizable surrogate models.
Specifically, we:
\begin{itemize}
    \item Utilize the novel modelfluid representation to create a surrogate model for distillation columns that generalizes over the entire space of homogeneous ternary vapor-liquid systems, making it truly reusable between different flowsheets, operating conditions, and chemical systems.
    \item Detail the generation of a large-scale training dataset using modern machine learning-based property prediction methods, supplemented with established pure component data from databases like DIPPR \cite{Wilding1998}.
    We generate a dataset of well-beyond $100,000$ individual mixtures and over $1,000,000$ unique data points.
    \item Successfully predict the distillation column outcomes for different ternary systems at different process specifications.
    % \item Discuss the challenge of ensuring thermodynamic consistency for these synthetically generated mixtures and outline the checks used to filter out unphysical systems.
    \item We demonstrate the trained surrogate model on an entrainer selection case study, where the same surrogate model is used for each of the three distillation columns in the flowsheet. Optimization based on the surrogate model successfully identifies the promising candidates from an entrainer pool.
\end{itemize}
The result is a powerful and reusable tool that significantly lowers the barrier for comprehensive process optimization and design space exploration.
By interconnecting it with other unit operation models to flowsheets, the surrogate model can be used in a variety of case studies for any ternary vapor-liquid system.

While the main focus is on the ternary distillation column, surrogate models for binary and ternary VLE as well as a binary distillation column are provided in the Supporting Information as a reference.

The remainder of this paper is organized as follows.
\sectionref{sec:literature-review} reviews the relevant literature on surrogate modeling in chemical engineering, positioning our work within the state-of-the-art.
In \sectionref{sec:methodology}, we detail the development of the surrogate model, from the feature and target selection to the ML-fueled dataset generation and model training.
\sectionref{sec:applications} presents a comprehensive case study on entrainer distillation, where we validate the model's performance and demonstrate its practical utility.
Finally, we conclude in \sectionref{sec:conclusion} with a summary of our findings and an outlook on future research directions.

\section{Literature Review}
\label{sec:literature-review}
While this review focuses on chemical process engineering, the fundamental challenges and strategies of surrogate modeling are common across many scientific and engineering disciplines.
The book by Forrester et al. \cite{Forrester2008} provides a comprehensive overview of surrogate modeling in engineering design, and its importance is continually highlighted in reviews on topics from materials science \cite{Peivaste2025} and the development of digital twins \cite{Barkanyi2021}.

\paragraph{Surrogate Model Types, Scope, and Sampling}
A variety of surrogate modeling methods are employed in the literature, with the choice often depending on the nature of the problem and the availability of data.
In the often small data regime of engineering, Bayesian methods such as Kriging or Gaussian Processes (GPs) are particularly popular \cite{Bishop2006}.
Due to their inherent ability to provide uncertainty estimates for their predictions, they are often used in combination with adaptive sampling strategies, where the goal is to iteratively acquire new data points to refine the model in specific regions of interest.
This approach, broadly known as Bayesian Optimization, is used for the training of local surrogate models whose primary purpose is to find an optimum within a defined design space, without needing global accuracy \cite{Caballero2008, Winz2025}.
Significant research has been contributed on this topic, including various adaptive sampling approaches for both Kriging- and Neural Network-based models \cite{Nentwich2019a,Nentwich2019b,Winz2021,Winz2024} and on feasibility analysis using such methods \cite{Dias2019}.
In contrast, the increasing availability of large datasets, sometimes generated via ML-based prediction methods, has fueled the rise of Artificial Neural Network (ANN) based surrogates.
ANNs are used for capturing complex, nonlinear relationships in large datasets, which enables the development of \textit{global} surrogate models that are accurate across vast feature spaces.
Such global models are the focus of this work, where training data is generated in a one-shot manner to create a pre-trained, reusable asset.

\paragraph{Levels of Abstraction in Process Engineering}
Surrogate models in chemical engineering are applied at various levels of abstraction, a choice that involves a critical trade-off between ease of use and model reusability.
\begin{itemize}
    \item \textbf{Flowsheet-level surrogates:} Traditionally, surrogates were often built for an entire process flowsheet \cite{Palmer2002a,Palmer2002b}.
    While straightforward to apply, these models are highly brittle: Any change in the flowsheet topology (e.g., adding a recycle stream) renders the surrogate obsolete.
    Furthermore, generating training data can be paradoxical, as it requires numerous successful simulations of the entire flowsheet, the very task that is often difficult and convergence-prone.
    \item \textbf{Unit operation-level surrogates:} A more robust approach, with historical roots in the modular nature of process simulation \cite{Freund2008}, is to develop surrogates for individual unit operations \cite{Caballero2008}.
    These modular surrogates can be interconnected to form different flowsheet configurations.
    Bubel et al. \cite{Bubel2021} demonstrated the advantages of this approach for a pressure-swing distillation process, showcasing improved flexibility and reusability over flowsheet-level models.
    This modularity is a core principle of the work presented here.
    \item \textbf{Property-level surrogates:} Descending to an even finer level of detail, researchers have developed surrogates for specific, computationally expensive calculations within unit models, such as phase equilibrium calculations based on the PC-SAFT equation of state \cite{Nentwich2019a, Winz2021} or the prediction of fugacity coefficients \cite{Nentwich2019b}.
\end{itemize}

\paragraph{A Unifying Limitation of the State-of-the-Art}
Despite the sophistication of these methods, a common limitation pervades the literature: the validity of a trained surrogate model is almost always restricted to a fixed chemical system.
Whether applied at the flowsheet, unit, or property level, the model is trained for specific components (e.g., water/ethanol).
If the chemical mixture changes, a completely new surrogate model must be generated, requiring a full cycle of data generation, training, and validation.
This is evident in recent works where numerous individual surrogates are required to study multiple systems \cite{Sethi2025, Abranches2023}, fundamentally limiting the scalability and industrial applicability of the surrogate modeling paradigm.

\paragraph{First Steps Towards Generalization Across Chemical Space}
To our knowledge, the work of Sun et al. \cite{Sun2023, Sun2024} represents the first significant attempt to create surrogate models that generalize across different chemical systems.
% checked it: they write the previous works tested some generalization ideas, in part, using similar descriptors to those mentioned in Sun2023, but they usually only consider a set of < 20 mixtures, which means that we can't speak of an actual "generalization" here.
They developed Artificial Neural Networks (ANN) and Graph Neural Network (GNN) based surrogates for binary vapor-liquid equilibria (VLE) prediction.
Their models take pure component properties (ANN) \cite{Sun2023} and molecular fingerprints (GNN) \cite{Sun2024} as inputs, respectively, alongside process conditions, to predict VLE behavior for a diverse set of chemical pairs.
This was a remarkable first step, demonstrating that a single model could generalize the effect of pure component descriptors on the vapor-liquid equilibria of binary mixtures over a decently large range of substances.
However, their work focused on showcasing GNN architectures for fundamental property prediction, rather than establishing a framework for reusable \textit{unit operation} models.
The feature vectors derived from GNNs are not directly interpretable and may not form a smooth manifold for optimization.
Crucially, their work addressed the challenge at the level of phase equilibrium, while the generalization of a complete, multi-variable unit operation model---with its complex interplay of mass and energy balances, physical constraints, and equipment parameters---remained an open challenge.
Moreover, while their dataset of 210 binary mixtures was the largest of its kind for this purpose, its scope remains limited, and generalization over the entire fluid mixture space is not guaranteed.

\paragraph{Positioning the Current Work}
This paper builds upon these foundations to address that very challenge.
We aim for a truly global surrogate model for a distillation column, intended to serve as a reusable prediction model for a unit operation.
The goal is to create a model that is pre-trained and can be deployed by users for a wide variety of ternary systems and applications that may not have been known at the time of training.
For this reason, our approach relies on an exhaustive one-shot sampling of the feature space to ensure global accuracy, rather than adaptive sampling targeting local fidelity.
While fine-tuning the model for specific, high-precision applications is possible and encouraged, this work focuses on establishing the core generalizability across the vast design space of column configurations and, for the first time, the chemical mixture space itself.

\section{Development of the Reusable Surrogate Model}
\label{sec:methodology}
This section details the development of the reusable surrogate model for distillation columns.
We first define the features and targets of the model, including the crucial modelfluid representation that enables generalization across chemical systems (\sectionref{sec:features}).
Next, we describe our novel ML-fueled dataset generation strategy, which allows us to create a large and diverse training dataset far exceeding the scale of previous works (\sectionref{sec:dataset}).
Finally, we outline the training procedure for the surrogate model, quantify its prediction uncertainty, and evaluate the achieved predictive performance (\sectionref{sec:training}).

We further note that the proposed surrogate model is reusable within the space of ternary mixtures obeying homogeneous vapor-liquid phase behavior.
While the modelfluid representation generally extends to multi-component systems, such an extension is beyond the scope of this work. We revisit this potential in \sectionref{sec:conclusion}, where we discuss future directions.

\subsection{Feature and Target Selection}
\label{sec:features}
The predictive power and generalizability of the surrogate model are fundamentally determined by the choice of its input features and output targets.
We designed these to capture both the thermodynamic behavior of the chemical mixture and the physical configuration of the unit operation.

\paragraph{The Modelfluid Representation}
The key enabler for generalization across the ternary mixture space is the modelfluid representation introduced in \cite{Bubel2025c}.
For the sake of a self-contained manuscript, we summarize its core principles here.
Instead of using discrete component identities (e.g., SMILES strings), the framework uses a vector of continuous, physically meaningful descriptors that characterize the thermodynamic behavior of a binary fluid system.
This allows the surrogate model to learn the underlying relationships between fluid properties and column performance, rather than memorizing the behavior of a few specific systems.
\newline
For a ternary system, the modelfluid is defined by the following features for each of the three components and their binary pairs:
\begin{itemize}
    \item \textbf{Pure Component Properties:} The saturated vapor temperatures ($\temperature_i^{\saturatedvapor}$) at a given system pressure and the corresponding vaporization enthalpies ($\vaporizationenthalpyof{i}$).
    These anchor the system's boiling behavior and energy requirements.
    \item \textbf{Mixture Interaction Properties:} The activity coefficients at infinite dilution ($\activitycoefficient_i\vert_j$) and the derivative of the vapor mole fraction with respect to the liquid mole fraction at infinite dilution ($\partial \vapormolarfractionof{i} / \partial \liquidmolarfractionof{i}\vert_j$).
    The former is a direct measure of liquid-phase non-ideality and is predictable via methods like matrix completion \cite{Jirasek2020, Damay2021}.
    The latter is a powerful, derivative-based feature that directly quantifies the relative volatility at infinite dilution, a key indicator of separation difficulty.
    For a detailed thermodynamic treatment of this feature, the reader is referred to \cite{Bubel2025c}.
\end{itemize}

We define the modelfluid feature vector as follows:
\begin{equation} \label{eq:modelfluid:features}
    \modelfluidfeatures = \left[\begin{aligned}
        &\pressure, \saturatedvaportemperatureof{1}, \saturatedvaportemperatureof{2}, \saturatedvaportemperatureof{3}, \vaporizationenthalpyof{1}\inb{\saturatedvaportemperatureof{1}}, \vaporizationenthalpyof{2}\inb{\saturatedvaportemperatureof{2}}, \vaporizationenthalpyof{3}\inb{\saturatedvaportemperatureof{3}},\\
        &\activitycoefficient_1\vert_2, \activitycoefficient_2\vert_1, \activitycoefficient_1\vert_3, \activitycoefficient_3\vert_1, \activitycoefficient_2\vert_3, \activitycoefficient_3\vert_2, \frac{\partial \vapormolarfractionof{1}}{\partial \liquidmolarfractionof{1}}\vert_3, \frac{\partial \vapormolarfractionof{2}}{\partial \liquidmolarfractionof{2}}\vert_1, \frac{\partial \vapormolarfractionof{3}}{\partial \liquidmolarfractionof{3}}\vert_1
    \end{aligned}\right].
\end{equation}

Note that in \eqref{eq:modelfluid:features}, we omit $\frac{\partial \vapormolarfractionof{1}}{\partial \liquidmolarfractionof{1}}\vert_2$, $\frac{\partial \vapormolarfractionof{2}}{\partial \liquidmolarfractionof{2}}\vert_3$, and $\frac{\partial \vapormolarfractionof{3}}{\partial \liquidmolarfractionof{3}}\vert_2$.
This is due to the requirement of unique vapor pressure models for each component in a ternary system.
% A critical consideration for multi-component systems when using our modelfluid representation is the uniqueness of the pure components.
% For a ternary system, the vapor pressure model for any given component must be unique, regardless of the binary pair it is considered in.
This physical requirement, which is derived in detail in \cite{Bubel2025c}, reduces the number of independent features from 19 to 16, as in \eqref{eq:modelfluid:features}.
While we could also have chosen another subset of features for \eqref{eq:modelfluid:features}, the choice of features in this work advocates the incorporation of prediction methods for the activity coefficient at infinite dilution.

In our modeling \eqref{eq:modelfluid:features}, the activity coefficients at infinite dilution directly parameterize a Margules activity coefficient model.
The remaining features are used to explicitly calculate the parameters of a simplified two-parameter Antoine equation for vapor pressure and a composition-weighted model for the enthalpy of vaporization.
This feature-to-parameter mapping is deterministic and explicit, and its full set of equations is featured in \cite{Bubel2025c}.

Furthermore, the features in \eqref{eq:modelfluid:features} generally occupy a bounded space; temperatures, pressures, and enthalpies are physically constrained, while the activity coefficients and their derivatives are observed to be empirically limited within the datasets considered.
A detailed proposal on modelfluid feature bounds is provided in \cite{Bubel2025c}.

\paragraph{Unit Operation Features}
To ensure the surrogate model is also generalizable across different column designs and operating conditions, the following features of the unit operation are included as inputs:
\begin{itemize}
    \item The number of equilibrium stages above and below the feed stage ($\mynum_{\stages}^{\belowfeed}$, $\mynum_{\stages}^{\abovefeed}$).
    \item The ratio of the bottom flow rate to the feed flow rate ($\spl$).
    \item The reflux ratio ($\refluxratio$).
\end{itemize}
The respective feature vector is defined as
\begin{equation} \label{eq:columnfeatures}
    \features^{\column} = \left[\spl, \refluxratio, \mynum_{\stages}^{\belowfeed}, \mynum_{\stages}^{\abovefeed}\right] ,
\end{equation}
while the pressure $\pressure$ is already included in the modelfluid features $\modelfluidfeatures$.
% This feature set represents a carefully considered choice based on experience, though alternative representations are possible.
The final combined feature vector serving as input for the surrogate model is thus defined as:
\begin{equation} \label{eq:combinedfeatures}
    \features = \left[\modelfluidfeatures, \features^{\column}\right] .
\end{equation}

\paragraph{Rigorous Distillation Column Simulation}
All training, validation, and test data points -- as described in the following section -- are generated using a rigorous stage-to-stage distillation column model.
Given a feature vector \eqref{eq:combinedfeatures}, we can simulate a distillation column based MESH equation-based modeling (we refer to this as \textit{rigorous}) presented in \cite{Bubel2025c}.
The features from \eqref{eq:combinedfeatures} that describe the modelfluid \eqref{eq:modelfluid:features} are mapped to thermodynamic model parameters using the explicit feature-to-parameter map, as discussed above.
The features that describe the column specifications can straightforwardly be used in the rigorous column modeling.
Furthermore, to use the modelfluid-specific variant of MESH equation-based distillation column simulation presented in \cite{Bubel2025c}, we need to make the following assumptions:
\begin{enumerate}
    \item Streams outside columns are at liquid-boiling state.
    \item Columns operate at a constant, uniform pressure.
    \item Columns are adiabatic, with heat exchange only at the condenser and reboiler.
    \item Total condensation and evaporation are assumed for the condenser and reboiler.
    \item The enthalpy of mixing is neglected.
\end{enumerate}
\usetikzlibrary{arrows.meta, positioning, calc}
\newcommand{\productStreamXLength}{1.5}
\newcommand{\productStreamYLength}{0.5}
\newcommand{\columnXLength}{1.5}
\newcommand{\columnYLength}{4}
\newcommand{\unitDistance}{2.5}
\newcommand{\mixerLength}{1}
\newcommand{\streamspec}[1]{$\liquidflow_{#1},\liquidmolarfractions_{#1}$}

\begin{figure}
    \centering
    \begin{tikzpicture}[auto]

        % Define styles for blocks and streams
        \tikzstyle{column} = [draw, rectangle, minimum width=\columnXLength cm, minimum height=\columnYLength cm, rounded corners=10pt]
        \tikzstyle{stream} = [->, thick]

        % Columns
        \node[column, align=center] (C1) at ($(
            \unitDistance,0)$) {$\mynum_{\stages}^{\abovefeed}$,\\$\mynum_{\stages}^{\belowfeed}$,\\$\pressure$};

        % Column 1 Streams
        \draw[stream] ($(-\unitDistance/3,0)$) -- node[above] {\streamspec{\feed}} (C1);
        \draw[stream] (C1.north) -- ++(0,\productStreamYLength*1.5) -- ++(\productStreamXLength,0) node[above, xshift=\productStreamXLength*0.5 cm] {\streamspec{\distillate}};
        \draw[stream] (C1.south) -- ++(0,-\productStreamYLength) -- ++(\unitDistance/2+\productStreamXLength/2,0) node[below, xshift=\productStreamXLength*0.5 cm] {\streamspec{\bottom}};

        % Recycle streams for Column 1
        \draw[stream] (C1.north) -- ++(0,\productStreamYLength*1.5) -- ++(\productStreamXLength*2/3,0) -- ++(0,-\productStreamYLength*1.5) -- ++(0,-\productStreamYLength) -- ++(-\productStreamXLength*2/3+\columnXLength/2,0);
        \draw[thick] (C1.north) -- ++(0,\productStreamYLength*1.5) -- ++(\productStreamXLength*2/3,0) node[above] {$\refluxratio$};
        \draw[stream] (C1.south) ++(0,-\productStreamYLength) -- ++(\productStreamXLength*2/3,0) -- ++(0,\productStreamYLength) node[right, xshift=\productStreamXLength, yshift=0.25 cm] {$\reboilerduty$} -- ++(0,\productStreamYLength) -- ++(-\productStreamXLength*2/3+\columnXLength/2,0);
        \draw[thick] (C1.south) -- ++(0,-\productStreamYLength) -- ++(\productStreamXLength*2/3,0) node[below] {$\boilupratio$};

        % Reboiler and Condenser for column 1
        \draw[thick] (C1.north) ++(0,\productStreamYLength*0.75) circle (0.2);
        \draw[thick] (C1.north) ++(0,\productStreamYLength*0.75) node[left, xshift=-0.1 cm] {$\condenserduty$};
        \draw[thick] (C1.south) ++(0,-\productStreamYLength) ++(\productStreamXLength*2/3,0) ++(0,\productStreamYLength) circle (0.2);

    \end{tikzpicture}
    
    \caption{
        Schematic of a distillation column used in this work.
        The column processes feed ($\feed$) into distillate ($\distillate$) and bottom ($\bottom$) streams, operating at pressure $\pressure$.
        Its size is defined by the number of stages above ($\mynum_{\stages}^{\abovefeed}$) and below ($\mynum_{\stages}^{\belowfeed}$) the feed stage.
        Heat duties at the reboiler and condenser are represented by $\reboilerduty$ and $\condenserduty$, respectively.
        The column's operation is characterized by the boilup ratio ($\boilupratio$) and reflux ratio ($\refluxratio$), respectively, which control how much of the bottom and distillate output is recycled back to the column.
        For simplicity, internal stages are not shown.
    }
    \label{fig:distillation-column}
\end{figure}
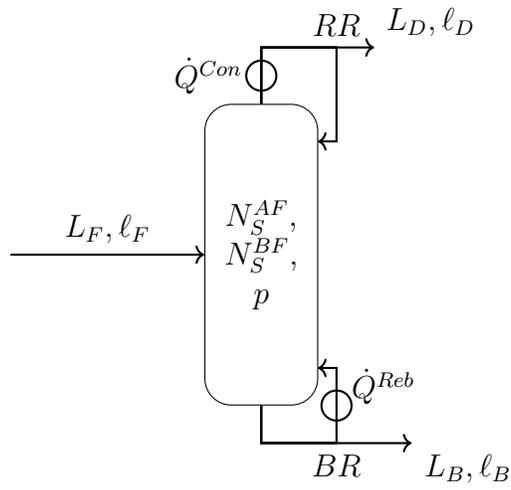

A sketch of the column, as considered in this work, is shown in \figureref{fig:distillation-column}.

\paragraph{Model Targets}
The surrogate model is trained as an explicit map to predict key performance indicators and separation outcomes.
Given the full feature vector \eqref{eq:combinedfeatures}, the model outputs the following targets:
\begin{itemize}
    \item The reboiler heat duty ($\reboilerduty$) (condenser duty can be added but is not considered in this work).
    \item The bottom stream molar fractions ($\liquidmolarfractions_{\bottom,i} \quad \forall i = 1, \dots, \mynum_{\component}-1$).
    \item The distillate stream molar fractions ($\liquidmolarfractions_{\distillate,i} \quad \forall i = 1, \dots, \mynum_{\component}-1$).
\end{itemize}
We only predict the molar fractions of $\mynum_{\component}-1$ components, as the last component's fraction is determined by the summation constraint
\begin{equation}
    \nonumber
    \liquidmolarfractionof{\mynum_{\component}} = 1 - \sum_{i=1}^{\mynum_{\component}-1} \liquidmolarfractionof{i}
    .
\end{equation}
While the prediction of both bottom and distillate compositions is redundant, we choose to predict both in order to analyze and compare the model's predictive accuracy on each stream.

\subsection{Dataset Generation}
\label{sec:dataset}
The generation of datasets for the development of a distillation column surrogate model includes both, the sampling of modelfluid features and the rigorous simulation of the distillation column for those modelfluid features at various operating conditions -- as described in \sectionref{sec:features}.
% Note that the term \textit{rigorous simulation} refers to the MESH equation-based distillation column modeling, using the modelfluid representation (two-parameter Antoine equation for vapor pressure, Margules activity coefficient model, and composition-weighted enthalpy of vaporization model -- as explained above) as the underlying thermodynamic model.
While the final surrogate model is trained to operate solely on the modelfluid features \eqref{eq:modelfluid:features}, the explicit map of modelfluid features to thermodynamic model parameters is the essential bridge that enables the use of rigorous column simulation for creating the large-scale dataset.

For the generation of training-, validation-, and test datasets in this work, we do not randomly sample modelfluid features.
Instead, we use both, information available from pure component property- and mixture property databases and combine them with ML-predicted mixture properties.
The below paragraphs, describe how the modelfluid features are obtained for the different datasets used to train, validate and test the surrogate model.

The generation of all datasets includes distillation column simulation for the modelfluid feature samples.
To prevent simulations at thermodynamically inconsistent modelfluid feature samples, which is generally possible, we perform a series of VLE-based consistency checks, as detailed in \appendixref{sec:vlechecks}.
If any of those checks fails for a modelfluid feature vector, it is discarded from the dataset.

For each modelfluid feature vector, we perform $100$ at random column specifications, to sample the column features \eqref{eq:columnfeatures}.
The column specifications are uniformly sampled from the following bounds:
\begin{itemize}
    \item $\liquidmolarfractions_{\feed} \in \left[0, 1\right]^3$ s.t. $\sum x_i = 1$
    \item $\refluxratio \in \left[0.1, 40\right]$
    \item Bottom-to-Feed flow rate split: $\spl \in \left[0.001, 0.999\right]$
    \item Number of stages above feed: $\mynum_{\stages}^{\abovefeed} \in \left[2, 30\right]$
    \item Number of stages below feed: $\mynum_{\stages}^{\belowfeed} \in \left[2, 30\right]$
    \item Pressure: $\pressure \in \left[0.5, 10\right]$ bar
\end{itemize}

\paragraph{ML-fueled Training and Validation Dataset}
A cornerstone of this work is the generation of a large-scale dataset sufficient for training a global surrogate model in a high-dimensional feature space.
Our approach represents a story of \textbf{ML-for-AI}: we leverage an existing machine learning method to generate a large dataset, which in turn enables the successful training of our artificial intelligence-based surrogate model.
What this means is that the data for training and validation is generated synthetically.
The modelfluid representation for each ternary mixture is constructed by combining pure component data with ML-predicted mixture properties.
Specifically, pure component Antoine equation parameters and the pure component vaporization enthalpies model parameters are sourced from the commercial DIPPR 801 database \cite{Wilding1998}.
Using those, we can compute the saturated vapor temperatures $\saturatedvaportemperatureof{i}$ and the pure component vaporization enthalpies at the saturation temperature $\molarenthalpy_i^{\saturatedvapor}$.
The crucial mixture interaction information is obtained by predicting the activity coefficients at infinite dilution ($\gamma^\infty$).
While many powerful prediction methods have recently been developed \cite{Rittig2023,Medina2023a,DiCaprio2023,Specht2024,Damay2021}, we use the Matrix Completion method from \cite{Jirasek2020}, as it has been successfully applied with this modelfluid representation in \cite{Bubel2025c}.
These $\activitycoefficientatinfinitedilutionin{i}{j}$ values parameterize the Margules model of the mixture.
Using the vapor pressure Antoine equations and the Margules activity coefficient model, we compute the $\derivativefeatureatinfdilution{i}{j}$ at a chosen pressure $\pressure$.
This yields all features of \eqref{eq:modelfluid:features}, which means that we can use the rigorous modelfluid-based distillation column simulation as discussed in \sectionref{sec:features}.

The power of this setup comes from the usage of the ML-based activity coefficient prediction.
We can compute the modelfluid feature vector for each ternary combination of components found in our pure component property database (which is vast), resulting in $167,221$ unique and consistent modelfluid feature vectors (excluding samples that fail the VLE checks \appendixref{sec:vlechecks}), i.e. $167,221$ unique ternary mixtures.
As a reference, the works of Sun et al. \cite{Sun2023, Sun2024} were based on 210 binary mixtures.

Performing distillation column simulation, as described above, we obtain a final dataset of $\num{2465106}$ feature and target vectors.

\paragraph{Test Dataset}
For final testing, a separate, higher-fidelity dataset was constructed using curated NRTL parameters from the \aspen~mixture property database \cite{AspenPlusV10}.
This means we combine the vapor pressure- and enthalpy models from the DIPPR 801 database \cite{Wilding1998} with the NRTL parameters from Aspen \cite{AspenPlusV10} and compute the vapor-liquid equilibria at some pressure $\pressure$.
From the obtained VLE, we determine the modelfluid features for 420 unique ternary mixtures, which form the basis of the test dataset. 
For each of these mixtures, we generate a range of distillation column specifications by sampling the relevant column features as described above.
This results in a comprehensive test dataset comprising $\num{13,494}$ distinct feature-target pairs, each representing a unique combination of mixture and column operating conditions.

This dataset, while smaller, is considered a suitable test of the surrogate model's predictive performance for real-world industrially-relevant systems, after being trained on synthetically generated data using ML-based predictions of limiting activity coefficients.

\paragraph{Discussion on Data Distribution}
\begin{figure}[h!]
    \centering
    \includegraphics[width=\textwidth]{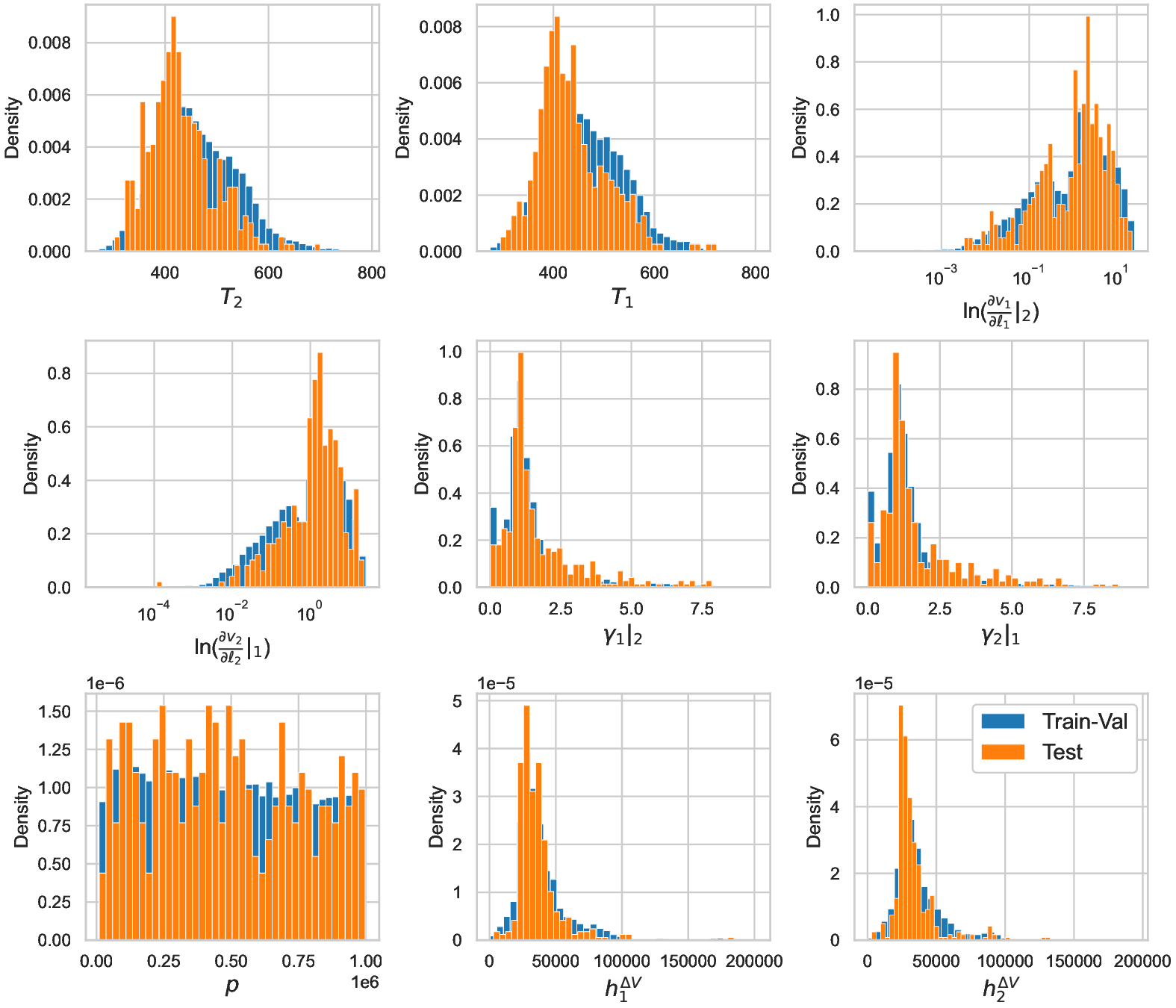}
    \caption{
        Distribution of selected modelfluid features in the training and validation dataset (blue) and the test dataset (orange), showing a reasonable overlap despite different data sources.
        The term \textit{density} on the vertical axes means that the vertical axis values (frequency) have been normalized such that the total area under the distribution curve sums to one.
    }
    \label{fig:feature_space}
\end{figure}
As illustrated in \figureref{fig:feature_space}, the sampling of the modelfluid feature space in our training data is not perfectly uniform.
This is a consequence of the natural distribution of the predicted activity coefficients and the filtering effect of our VLE consistency checks.
This non-uniformity appears to be beneficial compared to a uniform sampling of modelfluid features, as it better reflects the distribution of real-world chemical systems found in our test set.
This may also be due to undetected artifacts in the generated data, when sampling uniformly in the modelfluid feature space, which remains a challenge and motivates future work on more robust consistency checks for large-scale thermodynamic data generation.
% Our initial efforts using random uniform exploration of the feature space were less successful.
% For example, after applying the same VLE consistency checks (Appendix~\ref{sec:vlechecks}) to a uniformly sampled dataset, the remaining valid systems consisted of over 90\% maximum-boiling azeotropes.
% Such an ill-balanced dataset is poorly suited for training a general-purpose regression model, as it would be heavily biased towards one type of phase behavior.
% The ML-fueled dataset, being anchored to property distributions from known chemical structures, results in a far more balanced and physically representative set of systems.
Additional visualizations of the modelfluid feature space samples of the training-, validation- and test dataset, including binary projection scatter plots, are provided in the Supporting Information.

\subsection{Model Training and Analysis} \label{sec:modeltraining}
\label{sec:training}
With the features, targets, and datasets defined, the final step is the training and calibration of the surrogate model.
The model is an Artificial Neural Network (ANN), chosen for its ability to approximate highly complex, nonlinear functions.

We train a total of four individual ANN-based surrogate models in this work: a VLE prediction model for binary and ternary systems, and a distillation column surrogate model for binary and ternary systems.
In this section, we only discuss the distillation column surrogate model for ternary systems, as it is the most difficult to train and generalize.
Detailed descriptions of the model architectures, training hyperparameters, and performance evaluations for the remainder of the models are provided in the Supporting Information

\begin{figure}[h]
    \centering
    \includegraphics[width=\textwidth]{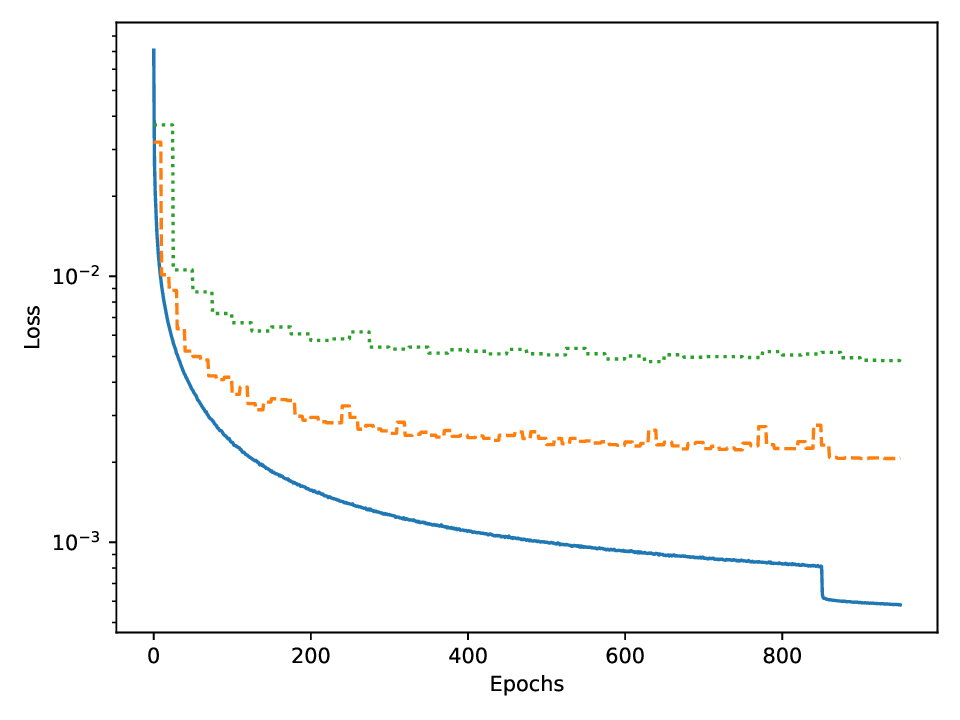}
    \caption{Evolution of the Mean Squared Error (MSE) loss on the training (blue solid line), validation (orange dashed line), and test (green dotted line) sets over the training epochs.}
    \label{fig:ternarycolumn:loss}
\end{figure}

The ML-generated dataset was split randomly into training (80\%) and validation (20\%) sets.
Even though overfitting was not observed to be an issue in this work, as confirmed by the learning curves shown in \figureref{fig:ternarycolumn:loss}, we manually excluded all ternary mixtures used in the entrainer selection case study \sectionref{sec:applications} from the training dataset, to provide a rigorous out-of-sample validation.
% For the specific application case studies presented later in this work (see \sectionref{sec:applications}), we ensured that the corresponding chemical systems were deliberately excluded from the training set.
% This provides a rigorous out-of-sample validation, even though general overfitting was not observed to be an issue, as confirmed by the learning curves shown in \figureref{fig:ternarycolumn:loss}.

We use a feed-forward neural network with linear layers and rectified linear unit (ReLU) activation functions on all but the final layer.
The model uses one input layer, four hidden layers, and one output layer, with (22, 1024), (1024, 512), (512, 256), (256, 128), (128, 64), (64, 5) neurons in each layer, where the first number is the number of input features and the second number is the number of output features of each layer, respectively.
We used a batch size of 64 and the model was trained for 800 epochs on a learning rate of 0.0001 and 200 epochs on a learning rate of $5e-5$.
Those settings were chosen based on previous experience, as well as a small number of experimental runs.
The training, validation, and test dataset consist of a total of \num{1972085}, \num{493041}, and \num{13494} data points, respectively.

\begin{figure}[h]
    \centering
    \includegraphics[width=\textwidth]{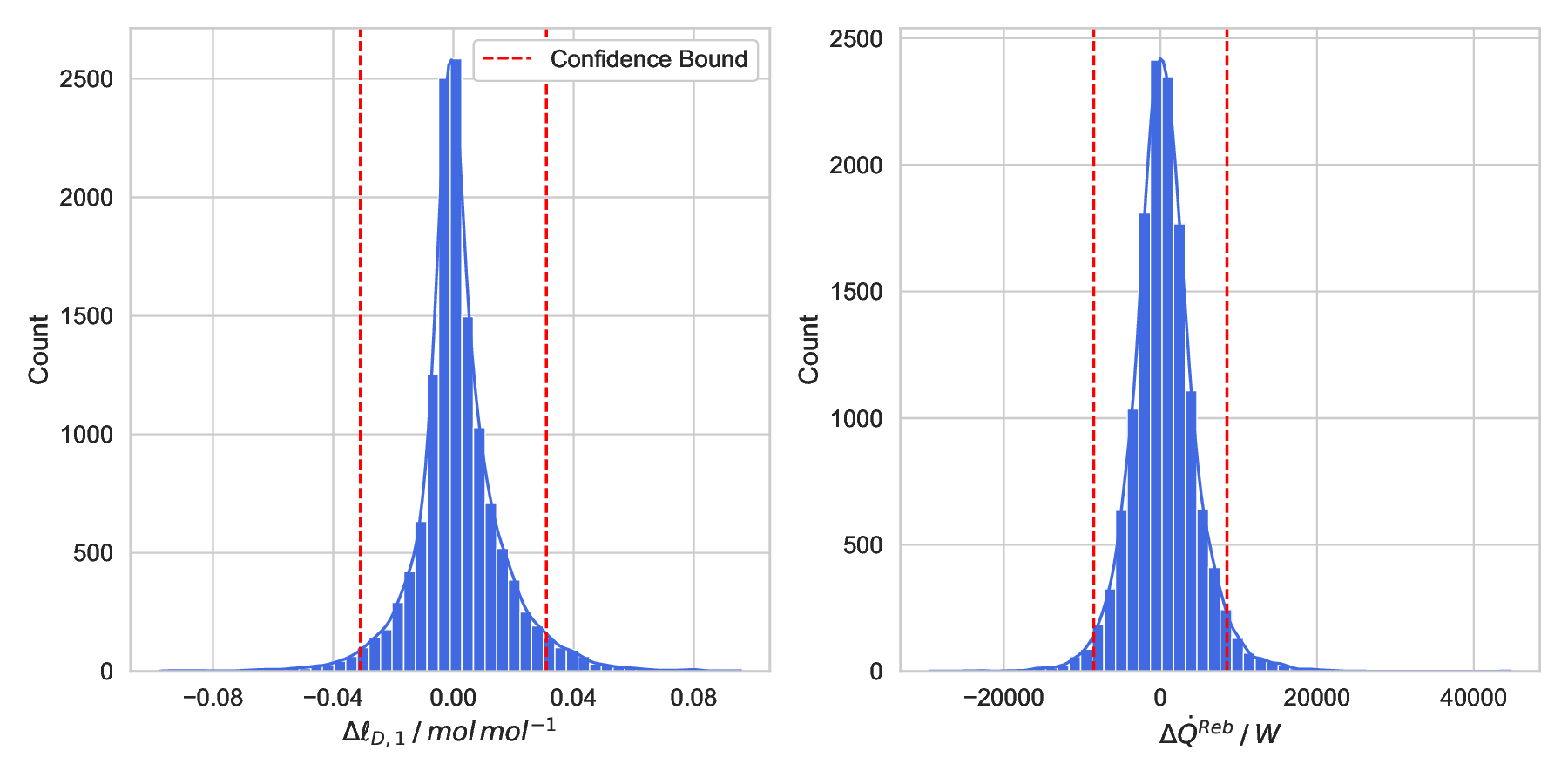}
    \caption{
        Histogram of prediction errors where $\Delta \liquidmolarfractionof{\distillate, 1} = \liquidmolarfractionof{\distillate, 1}^{\rigorous} - \liquidmolarfractionof{\distillate, 1}^{\surrogatemodel}$ and $\Delta \reboilerduty = \dot{Q}^{\reboiler,\rigorous} - \dot{Q}^{\reboiler,\surrogatemodel}$.
        The property with superscript $\rigorous$ denotes the result of rigorous distillation column simulation and the property with superscript $\surrogatemodel$ denotes the result of the surrogate model prediction.
        The red-dashed lines show the confidence regions obtained from calibrating the conformal prediction on the validation dataset.
        There are a couple (less than 20 overall) of outliers with $\Delta \liquidmolarfractionof{\distillate, 1}$ errors $> 0.1$, which we removed from the figure for the sake of axis scaling.
    }
    \label{fig:ternarycolumn:error_histogram}
\end{figure}

\paragraph{Performance Evaluation}
The predictive accuracy of the trained surrogate model is evaluated by comparing its predictions against the ground truth values from the test dataset.
The distribution of the prediction errors, shown in \figureref{fig:ternarycolumn:error_histogram}, indicates that the model achieves high accuracy for the vast majority of cases.

\begin{figure}[h]
    \centering
    \includegraphics[width=\textwidth]{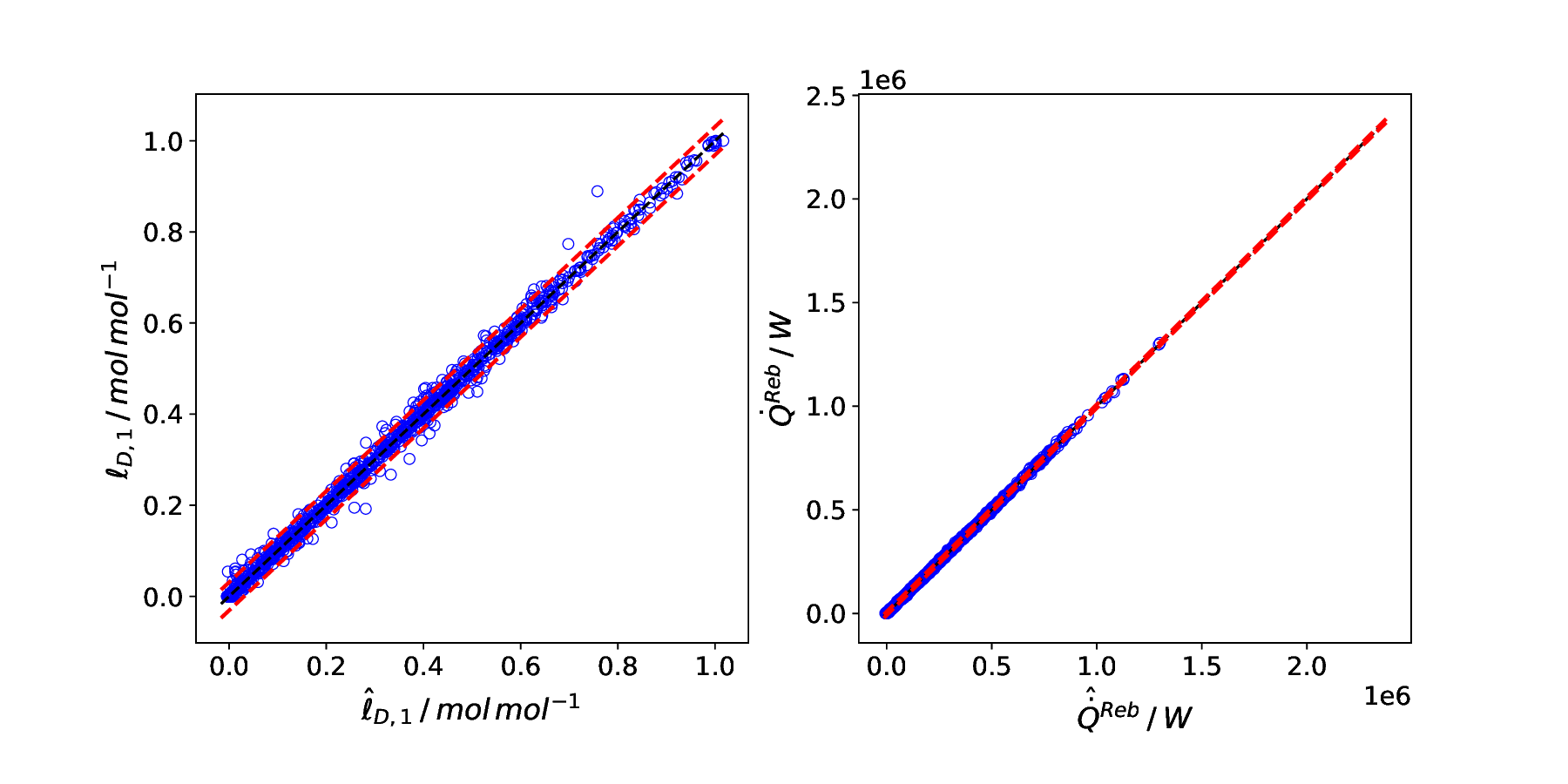}
    \caption{
        Measured versus predicted values for the ternary distillation column surrogate model on the test dataset.
        Surrogate model predictions (horizontal axis, denoted by a $\hat{\dot}$ symbol) are compared to ground truth values from rigorous simulation (vertical axis).
        The red dashed lines indicate the 95\% prediction intervals, calibrated via conformal prediction on the validation set.
        The close clustering along the diagonal demonstrates the model's high accuracy and generalizability across diverse ternary systems and column configurations.
        The figure displays a randomly selected subset of the test dataset, as visualizing the entire dataset is impractical due to its large size.
    }
    \label{fig:ternarycolumn:predicted_vs_measured}
\end{figure}

The results shown in \figureref{fig:ternarycolumn:error_histogram} and \figureref{fig:ternarycolumn:predicted_vs_measured} demonstrate that the trained surrogate model successfully generalizes across both the space of distillation column specifications and the space of ternary systems with homogeneous vapor-liquid phase behavior.
This makes the trained surrogate model \textbf{reusable}, as it can be used for various systems and operating conditions, without the necessity of re-calibration or fine-tuning.
For brevity, error histograms and measured vs. predicted diagrams for all molar fraction targets are not presented, as their distributions are very similar.
The Root Mean Square Error (RMSE) for all targets, summarized in \tabref{tab:rmse_values}, confirms this observation and quantifies the overall model performance.
\begin{table}[h!]
    \centering
    \begin{tabular}{|l|c|}
        \hline
        \textbf{Property} & \textbf{RMSE} \\
        \hline
        $\liquidmolarfractionof{\bottom, 1}$ & $0.0257 \: \molefracunit$ \\
        $\liquidmolarfractionof{\distillate, 1}$ & $0.0271 \: \molefracunit$ \\
        $\liquidmolarfractionof{\bottom, 2}$ & $0.0189 \: \molefracunit$ \\
        $\liquidmolarfractionof{\distillate, 2}$ & $0.0193 \: \molefracunit$ \\
        $\reboilerduty$ & $4.216 \: \text{k}\wattunit$ \\
        \hline
    \end{tabular}
    \caption{Root Mean Square Error (RMSE) values for the surrogate model predictions.}
    \label{tab:rmse_values}
\end{table}

According to \figureref{fig:ternarycolumn:predicted_vs_measured}, the model predicts the reboiler duty with higher accuracy than the bottom and distillate compositions.
This is because the reboiler duty's dependency on the surrogate model input features is largely monotonic.
For example, a lower bottom-to-feed split results in a higher reboiler duty.
This behavior is easier for the model to learn and generalize compared to the output concentration values, which depend on the complex, often non-monotonic azeotropic behavior of the system.
%
% We note that higher prediction accuracy for purely azeotropic systems could be achieved by using an alternative modeling approach that explicitly includes the location of the azeotropic point as input features, which simplifies the input-output mapping.
% However, such a specialized model would be limited to azeotropic mixtures and would not fulfill the primary goal of this work: to create a single, pre-trained surrogate that is reusable across the entire spectrum of zeotropic and azeotropic systems.
% A detailed discussion of this alternative modeling approach is provided in \appendixref{sec:app-alternativemodelfluidrepresentation}.

\paragraph{Confidence Interval Calibration via Conformal Prediction}
To provide a reliable measure of the model's prediction uncertainty, we employ the method of conformal prediction \cite{Angelopoulos2023}.
This post-processing technique is particularly well-suited for this application as it uses the validation dataset to calibrate prediction intervals with a statistically rigorous guarantee on marginal coverage.
This means that, for a chosen confidence level of 95\%, we can expect 95\% of the true values to fall within their respective predicted intervals over the long run, adding a crucial layer of trust and reliability to the surrogate model's outputs.

The method yields a symmetric 95\% confidence interval (CI) for each prediction $\hat{y}$ of the form:
\begin{equation} \label{eq:conformal_interval}
    \confidenceinterval = \left[ \estimator{\target} - \Delta, \estimator{\target} + \Delta \right]
\end{equation}
where $\Delta$ is the confidence interval radius calibrated on the validation set.
The calibrated radii for each of the model's output targets are presented in \tabref{tab:confidence_intervals}.

\begin{table}[h!]
    \centering
    \begin{tabular}{|l|c|}
        \hline
        \textbf{Property} & \textbf{Value} \\
        \hline
        $\Delta \liquidmolarfractions_{\bottom, \component_1}$ & $0.0355 \: \molefracunit$ \\
        $\Delta \liquidmolarfractions_{\bottom, \component_2}$ & $0.0355 \: \molefracunit$ \\
        $\Delta \liquidmolarfractions_{\distillate, \component_1}$ & $0.031 \: \molefracunit$ \\
        $\Delta \liquidmolarfractions_{\distillate, \component_2}$ & $0.0355 \: \molefracunit$ \\
        $\Delta \reboilerduty$ & $8.500 \: \text{k}\wattunit$ \\
        \hline
    \end{tabular}
    \caption{Confidence interval radii for surrogate model predictions.}
    \label{tab:confidence_intervals}
\end{table}

These radii, which correspond to the red-dashed lines shown in the error histogram in \figureref{fig:ternarycolumn:error_histogram}, quantify the model's expected predictive uncertainty.
For instance, the model predicts molar fractions with a typical uncertainty of approximately $\pm 0.035 \: \molefracunit$.

Eventually, it is the availability of an uncertainty quantification of the trained surrogate model that qualifies its use in practical applications, e.g. optimization case studies, where it is mandatory to estimate the uncertainty contained in the surrogate model-based predictions.
Now, the trained surrogate model can be interconnected to flowsheets and used in process simulation and optimization to support rigorous optimization and decision-making processes.
In the following section, we demonstrate this by applying the surrogate model to the problem of entrainer selection for an extractive distillation.

\section{Case Study: Surrogate Model-Based Entrainer Selection}
\label{sec:applications}
In this section, we demonstrate the practical application of the generalizable surrogate model by applying it to the problem of entrainer selection for extractive distillation.
The goal is to efficiently screen a pool of candidate entrainers for the separation of a maximum-boiling azeotropic mixture of Acetone and Chloroform.
This case study was also analyzed using a rigorous modeling approach in \cite{Bubel2025c}, providing a direct benchmark against which we can compare the performance and computational advantages of our surrogate-based methodology.

\paragraph{The Entrainer Distillation Process and Optimization Problem}
Entrainer distillation is a process that enhances the separation of azeotropic mixtures by introducing a third component, the entrainer, which favorably alters the vapor-liquid equilibrium \cite{Duessel1995}.
A schematic of the three-column process used in this work is shown in \figureref{fig:entrainer-distillation-flowsheet}.
% \usetikzlibrary{arrows.meta, positioning, calc}

% \newcommand{\productStreamXLength}{1.5}
% \newcommand{\productStreamYLength}{0.5}
% \newcommand{\columnXLength}{1.5}
% \newcommand{\columnYLength}{4}
% \newcommand{\unitDistance}{2.5}
% \newcommand{\mixerLength}{1}
\newcommand{\verticalstreamspec}[1]{%
$\begin{array}{c}
    \liquidflow_{#1} \\
    \liquidmolarfractions_{#1}
\end{array}$}
\newcommand{\horizontalstreamspec}[1]{$\liquidflow_{#1},\liquidmolarfractions_{#1}$}

\begin{figure}
    \centering
    \begin{adjustbox}{scale=0.8}
    \begin{tikzpicture}[auto]

        % Define styles for blocks and streams
        \tikzstyle{column} = [draw, rectangle, minimum width=\columnXLength cm, minimum height=\columnYLength cm, rounded corners=10pt]
        \tikzstyle{mixer} = [draw, rectangle, minimum size=\mixerLength cm]
        \tikzstyle{stream} = [->, thick]

        % Mixer
        \node[mixer] (M) at (0,0) {$M$};

        % Columns
        \node[column] (C1) at ($(M.east) + (\unitDistance,0)$) {$\column_1$};
        \node[column] (C2) at ($(C1.east) + (\unitDistance,0)$) {$\column_2$};
        \node[column] (C3) at ($(C2.east) + (\unitDistance,0)$) {$\column_3$};

        % Streams
        \draw[stream] (-\productStreamXLength,0) node[left] {\verticalstreamspec{\feed}} -- (M);
        \draw[stream] (M) -- node[above] {\verticalstreamspec{\feed\column_1}} (C1);

        % Column 1 Streams
        \draw[stream] (C1.north) -- ++(0,\productStreamYLength*1.5) -- ++(\productStreamXLength,0) node[above, xshift=\productStreamXLength*0.35 cm] {\verticalstreamspec{\product\column_1}};
        \draw[stream] (C1.south) -- ++(0,-\productStreamYLength) -- ++(\unitDistance/2+\productStreamXLength/2+0.1,0) -- ++(0,\productStreamYLength+\columnYLength/2) node[above, xshift=-3mm] {\verticalstreamspec{\recycle\column_1}} --++ (\unitDistance/2-\productStreamXLength/2-0.1,0);

        % Recycle streams for Column 1
        \draw[stream] (C1.north) -- ++(0,\productStreamYLength*1.5) -- ++(\productStreamXLength*2/3,0) -- ++(0,-\productStreamYLength*1.5) -- ++(0,-\productStreamYLength) -- ++(-\productStreamXLength*2/3+\columnXLength/2,0);
        \draw[thick] (C1.north) -- ++(0,\productStreamYLength*1.5) -- ++(\productStreamXLength*2/3,0) node[above, xshift=-3mm] {$\refluxratio_{\column_1}$};
        \draw[stream] (C1.south) ++(0,-\productStreamYLength) -- ++(\productStreamXLength*2/3,0) -- ++(0,\productStreamYLength) node[right, xshift=\productStreamXLength, yshift=0.25 cm] {$\reboilerduty_{\column_1}$} -- ++(0,\productStreamYLength) -- ++(-\productStreamXLength*2/3+\columnXLength/2,0);
        \draw[thick] (C1.south) -- ++(0,-\productStreamYLength) -- ++(\productStreamXLength*2/3,0) node[below] {$\boilupratio_{\column_1}$};

        % Reboiler and Condenser for column 1
        \draw[thick] (C1.north) ++(0,\productStreamYLength*0.75) circle (0.2);
        \draw[thick] (C1.north) ++(0,\productStreamYLength*0.75) node[left, xshift=-0.1 cm] {$\condenserduty_{\column_1}$};
        \draw[thick] (C1.south) ++(0,-\productStreamYLength) ++(\productStreamXLength*2/3,0) ++(0,\productStreamYLength) circle (0.2);

        % Column 2 Streams
        \draw[stream] (C2.north) -- ++(0,\productStreamYLength*1.5) -- ++(\unitDistance/2+\productStreamXLength/2,0) -- ++(0,-\productStreamYLength-\columnYLength/2) node[below, xshift=-2mm] {\verticalstreamspec{\recycle\column_2}} --++ (\unitDistance/2-\productStreamXLength/2,0);
        \draw[stream] (C2.south) -- ++(0,-\productStreamYLength) -- ++(\productStreamXLength,0) node[right, xshift=-3mm] {\verticalstreamspec{\product\column_2}};

        % Recycle streams for Column 2
        \draw[stream] (C2.north) ++(0,\productStreamYLength*1.5) -- ++(\productStreamXLength*2/3,0) -- ++(0,-\productStreamYLength*1.5) -- ++(0,-\productStreamYLength) -- ++(-\productStreamXLength*2/3+\columnXLength/2,0);
        \draw[thick] (C2.north) -- ++(0,\productStreamYLength*1.5) -- ++(\productStreamXLength*2/3,0) node[above, xshift=-1mm] {$\refluxratio_{\column_2}$};
        \draw[stream] (C2.south) ++(0,-\productStreamYLength) -- ++(\productStreamXLength*2/3,0) -- ++(0,\productStreamYLength) node[right, xshift=\productStreamXLength, yshift=0.25 cm] {$\reboilerduty_{\column_2}$} -- ++(0,\productStreamYLength) -- ++(-\productStreamXLength*2/3+\columnXLength/2,0);
        \draw[thick] (C2.south) -- ++(0,-\productStreamYLength) -- ++(\productStreamXLength*2/3,0) node[below] {$\boilupratio_{\column_2}$};

        % Reboiler and Condenser for column 2
        \draw[thick] (C2.north) ++(0,\productStreamYLength*0.75) circle (0.2);
        \draw[thick] (C2.north) ++(0,\productStreamYLength*0.75) node[left, xshift=-0.1 cm] {$\condenserduty_{\column_2}$};
        \draw[thick] (C2.south) ++(0,-\productStreamYLength) ++(\productStreamXLength*2/3,0) ++(0,\productStreamYLength) circle (0.2);

        % Column 3 Streams
        \draw[stream] (C3.north) -- ++(0,\productStreamYLength*1.5) -- ++(\productStreamXLength,0) node[right, xshift=-2mm, yshift=2mm] {\verticalstreamspec{\product\column_3}};
        \draw[stream] (C3.south) -- ++(0,-\productStreamYLength) -- ++(\productStreamXLength*2/3,0) -- ++(0,-\productStreamYLength*1.5) -- ++(-\productStreamXLength*2/3-\unitDistance*3-\columnXLength-\mixerLength/2,0) node[below] {\verticalstreamspec{\recycle\column_3}} -- ++(0,\productStreamYLength*2.5+\columnYLength/2-\mixerLength/2);

        % Recycle streams for Column 3
        \draw[stream] (C3.north) ++(0,\productStreamYLength*1.5) -- ++(\productStreamXLength*2/3,0) -- ++(0,-\productStreamYLength*1.5) -- ++(0,-\productStreamYLength) -- ++(-\productStreamXLength*2/3+\columnXLength/2,0);
        \draw[thick] (C3.north) -- ++(0,\productStreamYLength*1.5) -- ++(\productStreamXLength*2/3,0) node[above, xshift=-2mm] {$\refluxratio_{\column_3}$};
        \draw[stream] (C3.south) ++(0,-\productStreamYLength) -- ++(\productStreamXLength*2/3,0) -- ++(0,\productStreamYLength) node[right, xshift=\productStreamXLength, yshift=0.25 cm] {$\reboilerduty_{\column_3}$} -- ++(0,\productStreamYLength) -- ++(-\productStreamXLength*2/3+\columnXLength/2,0);
        \draw[thick] (C3.south) -- ++(0,-\productStreamYLength) -- ++(\productStreamXLength*1/3,0) node[right, xshift=5mm, yshift=-3mm] {$\boilupratio_{\column_3}$};

        % Reboiler and Condenser for column 3
        \draw[thick] (C3.north) ++(0,\productStreamYLength*0.75) circle (0.2);
        \draw[thick] (C3.north) ++(0,\productStreamYLength*0.75) node[left, xshift=-0.1 cm] {$\condenserduty_{\column_3}$};
        \draw[thick] (C3.south) ++(0,-\productStreamYLength) ++(\productStreamXLength*2/3,0) ++(0,\productStreamYLength) circle (0.2);

    \end{tikzpicture}
    \end{adjustbox}
    \caption{
        Entrainer distillation flowsheet for the separation of a binary mixture with maximum-boiling azeotropic phase behavior.
        The mixing unit is denoted by $M$, and the distillation columns are labeled as $\column_i$.
        \horizontalstreamspec{\feed} denotes the feed stream flowrate and molar fractions.
        \horizontalstreamspec{\product\column_i} and \horizontalstreamspec{\recycle\column_i} represent the product and recycle stream flowrates and molar fractions from column $i$, respectively.
        \horizontalstreamspec{\feed\column_i} is the feed flowrate and molar fractions to column $i$.
        Due to the interconnection of columns in the flowsheet, we use the \product (product) and \recycle (recycle) flags instead of the traditional \bottom (bottom) and \distillate (distillate) stream labels.
        \horizontalstreamspec{\feed} denotes the flowsheet's feed stream flowrate and molar fractions.
    }
    \label{fig:entrainer-distillation-flowsheet}
\end{figure}
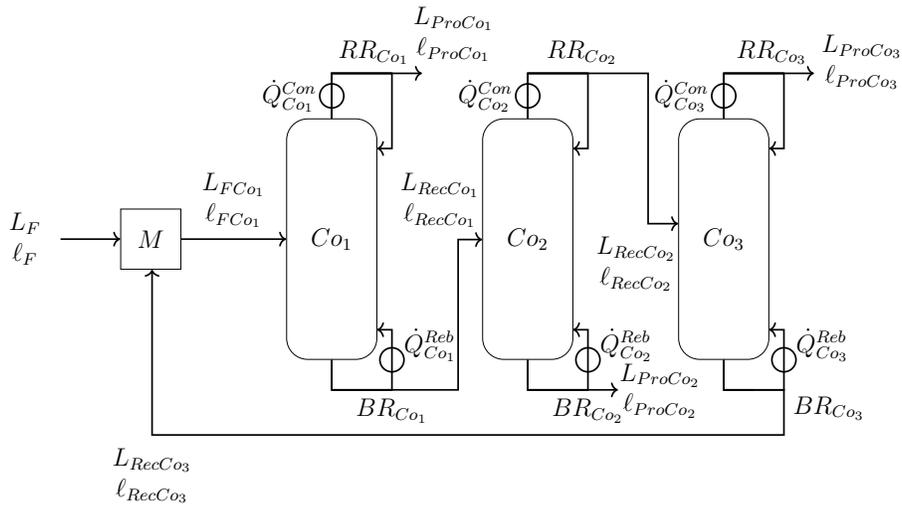

A common challenge in process design is balancing capital expenditures (CAPEX), represented by the column sizes, against operational expenditures (OPEX), represented by the energy consumption.
The selection of an appropriate entrainer plays a vital role in determining this balance, as it directly influences both the required number of stages and the necessary heat duty.
We address this by solving a multi-objective optimization problem to find the Pareto-optimal trade-off between minimizing the total number of stages, $\mynum_{\stages}^{\total}$, and the total reboiler heat duty, $\totalreboilerduty$.
The objective functions are:
\begin{equation} \label{eq:entrainerdistillation:objectivefunction:totalnumberofstages}
    \mynum_{\stages}^{\total} = \sum_{i=1}^{3} \mynum_{\stages, \column_i}
\end{equation}
\begin{equation} \label{eq:entrainerdistillation:objectivefunction:totalheatduty}
    \totalreboilerduty = \sum_{i=1}^{3} \reboilerduty_{\column_i}
\end{equation}
The optimization is subject to a set of rigorous constraints to ensure a physically meaningful and feasible process.
The product constraints require minimum purities \eqref{eq:entrainerdistillation:productpurity} and molar flow rates \eqref{eq:entrainerdistillation:productflow} for the product streams.
\begin{align} \label{eq:entrainerdistillation:productpurity}
    \liquidmolarfraction_{\product\component\column_i} \geq& \liquidmolarfraction_{\product\component\column_i}^{min} \quad &\forall i = 1, \dots, \mynum_{\column} \\
    \label{eq:entrainerdistillation:productflow}
    \liquidflow_{\product\column_i} \geq& \liquidflow_{\product\column_i}^{min} \quad &\forall i = 1, \dots, \mynum_{\column}
\end{align}
Furthermore, flowsheet mass balances must be satisfied, including stream closure conditions \eqref{eq:entrainerdistillation:closureconditions}, the overall component mass balance \eqref{eq:entrainerdistillation:flowsheetmassbalance}, and individual column mass balances \eqref{eq:entrainerdistillation:columnmassbalance}.
\begin{align} \label{eq:entrainerdistillation:closureconditions}
    1 - \sum_{j=1}^{\mynum_{\component}} \liquidmolarfraction_{\product\component_j\column_i} =& 0 & \forall i = 1, \dots, \mynum_{\column} \\ \nonumber
    1 - \sum_{j=1}^{\mynum_{\component}} \liquidmolarfraction_{\recycle\component_j\column_i} =& 0 & \forall i = 1, \dots, \mynum_{\column}
\end{align}
\begin{equation}  \label{eq:entrainerdistillation:flowsheetmassbalance}
    \liquidflow_{\feed} \cdot \liquidmolarfraction_{\feed\component_j} - \sum_{i=1}^{\mynum_{\component}} \liquidflow_{\product\column_i} \cdot \liquidmolarfraction_{\product\component_j\column_i} = 0 \quad \forall j = 1, \dots, \mynum_{\component}
\end{equation}
\begin{align} \label{eq:entrainerdistillation:columnmassbalance}
    &\liquidflow_{\feed\column_i} \cdot \liquidmolarfraction_{\feed\component_j\column_i} - \liquidflow_{\product\column_i} \cdot \liquidmolarfraction_{\product\component_j\column_i} - \liquidflow_{\recycle\column_i} \cdot \liquidmolarfraction_{\recycle\component_j\column_i} = 0 \\ \nonumber
    &\forall i = 1, \dots, \mynum_{\column}, j = 1, \dots, \mynum_{\component}
\end{align}
Finally, the MESH equations governing the stage-to-stage thermodynamic equilibrium within each distillation column, \cite[see p. 226]{Biegler1997}, must be satisfied \eqref{eq:entrainerdistillation:meshconstraint}.
\begin{equation} \label{eq:entrainerdistillation:meshconstraint}
    \constraint_{\column}\inb{\dots} = 0
\end{equation}
To generate the Pareto frontier (the NQ curve), we solve a series of single-objective, epsilon-constraint problems, which is suitable for non-convex frontiers.
The scalarized problem is formulated as an MINLP, minimizing the total reboiler duty subject to an upper bound on the total number of stages, $\epsilon_{\mynum_{\stages}^{\total}}$:
\begin{equation} \label{eq:entraineroptimizationproblem}
    \min_{\feature_{\process}} \totalreboilerduty \quad \text{s.t.} \quad \mynum_{\stages}^{\total} \leq \epsilon_{\mynum_{\stages}^{\total}} \quad \text{and constraints } \eqref{eq:entrainerdistillation:productpurity}-\eqref{eq:entrainerdistillation:meshconstraint}.
\end{equation}

\paragraph{Modeling the Flowsheet: Rigorous vs. Surrogate-Based Approach}
The core difference between the two approaches lies in how the distillation columns and the overall flowsheet are modeled.
In the \textbf{rigorous approach}, each distillation column is modeled based on the full set of MESH equations \eqref{eq:entrainerdistillation:meshconstraint} as complex, non-convex equality constraints within the optimization problem.
Solving these constraints is computationally intensive and can fail to converge.

In the \textbf{surrogate-based approach}, the explicit nature of our model fundamentally changes the problem.
Instead of solving a large system of equations \eqref{eq:entrainerdistillation:meshconstraint}, the column's behavior is evaluated with a single forward pass of the neural network.
This reduces the computational cost and improves the robustness of the optimization.
To solve the flowsheet mass balances with interconnected surrogates, we chose a simultaneous approach.
The product compositions of each column are treated as optimization variables, and the mass balances across each unit operation are enforced as equality constraints.
This method prevents the accumulation of individual surrogate model errors that can occur in a sequential evaluation scheme \cite{Bubel2021}.

\subsection{Results: Entrainer Performance and Ranking} \label{sec:ranking}
The primary advantage of the surrogate-based approach is the ability to rapidly compute the entire NQ curve for every entrainer candidate in our pool---a task that is computationally expensive using the rigorous model.
% We present a comparison of the NQ curves for three representative entrainers below.

\begin{figure}[h!]
    \centering
    \includegraphics[width=\textwidth]{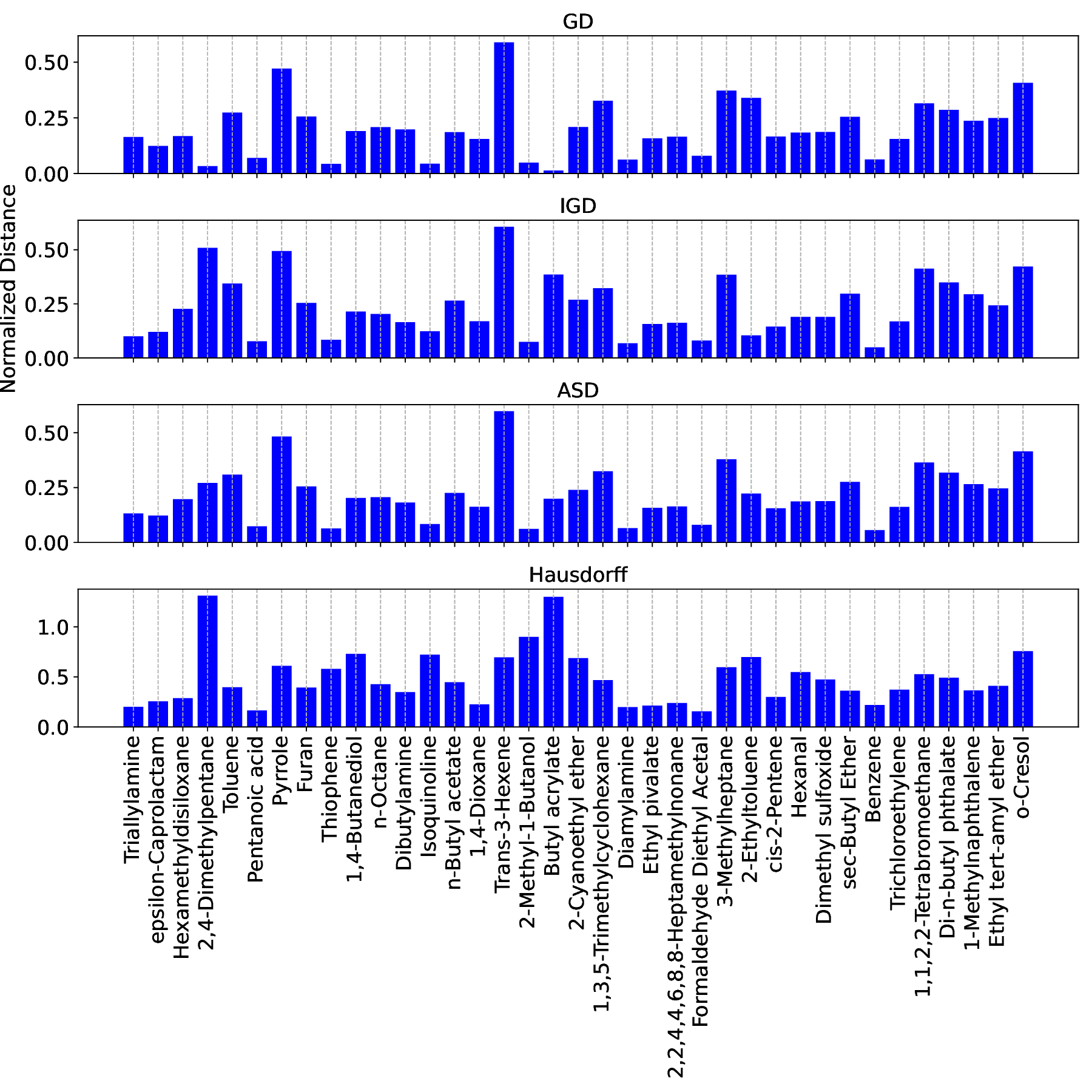}
    \caption{
        Histogram of distance metrics obtained from comparing the surrogate model-based NQ curves to those obtained from rigorous optimization.
        The component names on the horizontal axis correspond to the entrainer candidates used in the case study.
        The distance metrics correspond to the \textit{generational distance} \eqref{eq:gd}, \textit{inverted generational distance} \eqref{eq:igd}, \textit{average symmetric distance} \eqref{eq:asd}, and \textit{Hausdorff distance} \eqref{eq:hausdorff} and are defined in \appendixref{sec:distance_metrices}.
    }
    \label{fig:nq_curves_sats}
\end{figure}

To quantify the accuracy of the NQ curve predictions by the surrogate model, we compute and visualize distance measures for each entrainer candidate, c.f. \figureref{fig:nq_curves_sats}.
Since the quantification of distance measures between Pareto frontiers is non-trivial \cite{Ishibuchi2015}, we use multiple metrics to better quantify the results.
We use the generational distance (GD), inverted generational distance (IGD), average symmetric distance (ASD), and Hausdorff distance measures for the comparison in this section.
While motivated and explained in more detail in \appendixref{sec:distance_metrices}, a brief explanation of these metrics is:
\begin{itemize}
    \item GD: Measures the average distance of the surrogate model-based Pareto points to their closest neighbor on the rigorous Pareto frontier.
    \item IGD: Measures the average distance of the rigorous Pareto points to their closest neighbor on the surrogate model-based frontier.
    \item ASD: The mean value of GD and IGD.
    \item Hausdorff: Measures the maximum distance of a point on one frontier to its closest neighbor on the other frontier.
\end{itemize}
% We use the generational distance (GD) and inverted generational distance (IGD) measures, as described in \cite{Ishibuchi2015}, to assess the proximity of the surrogate model-based frontiers to that of the rigorous optimization and vice versa.
% We also use the average symmetric distance (ASD), which is the mean value of GD and IGD, and the Hausdorff distance to provide a more comprehensive comparison.
For all of these distance measures, a lower value indicates a better approximation of the true Pareto frontier.
% The GD metric \eqref{eq:gd}, which measures the average distance of the Pareto points of the surrogate model-based optimization to their respective closest-neighbor on the rigorous Pareto frontier, may not reveal if the surrogate model-based frontier does not cover the entire frontier from the rigorous optimization.
% On the other hand, the IGD metric \eqref{eq:igd}, which measures the average distance of the Pareto points of the rigorous optimization to their respective closest-neighbor on the surrogate model-based frontier, may not reveal if the surrogate model-based frontier contains any points that are far away from the rigorous Pareto frontier.
% This is why we also use the average symmetric distance (ASD) \eqref{eq:asd}, which yields the average of both the GD and IGD metrics.
% In addition, we use the Hausdorff distance metric \eqref{eq:hausdorff}, which measures the maximum distance of a point on one frontier to its closest-neighbor on the other frontier.
% This metric can be used as a warning flag, indicating when there are outliers on any of the curves in the comparison, both for the surrogate model-based and the rigorous optimization.
The overall observation from \figureref{fig:nq_curves_sats} is that most entrainer candidate's frontiers are successfully approximated by the surrogate, i.e. exhibit low distance metric values.
The results shown in \figureref{fig:nq_curves_sats} show the surrogate model's ability to generalize across different operating conditions and systems.
For some systems, the distance measures displayed in \figureref{fig:nq_curves_sats} exhibit significant variation, e.g. a small GD value but a large Hausdorff distance for Butyl acrylate.
% For example, the entrainer Butyl acrylate shows a large Hausdorff distance in \figureref{fig:nq_curves_sats}, while the GD value is small.
This indicates the existence of at least one point on the rigorous frontier that is far away from the surrogate model-based frontier.
Using the NQ curves for all entrainer candidates, provided in the Supporting Information to this work, this observation is confirmed.

% For Benzene (\figureref{fig:nq_curve_comparison_71-43-2}), a common entrainer, the surrogate model's predictions closely approximate the Pareto frontier obtained from the rigorous optimization.
% This demonstrates the model's ability to accurately capture the process behavior for well-behaved systems.

% \begin{figure}[h!]
%     \centering
%     \includegraphics[width=0.95\textwidth]{figures/nq_curve_comparison_71-43-2.eps}
%     \caption{
%         NQ curve comparison for Benzene (CAS 71-43-2) as entrainer in the system Acetone+Benzene+Chloroform.
%         The surrogate model provides a good approximation of the rigorous Pareto frontier.
%     }
%     \label{fig:nq_curve_comparison_71-43-2}
% \end{figure}

\begin{figure}[h!]
    \centering
    \includegraphics[width=0.95\textwidth]{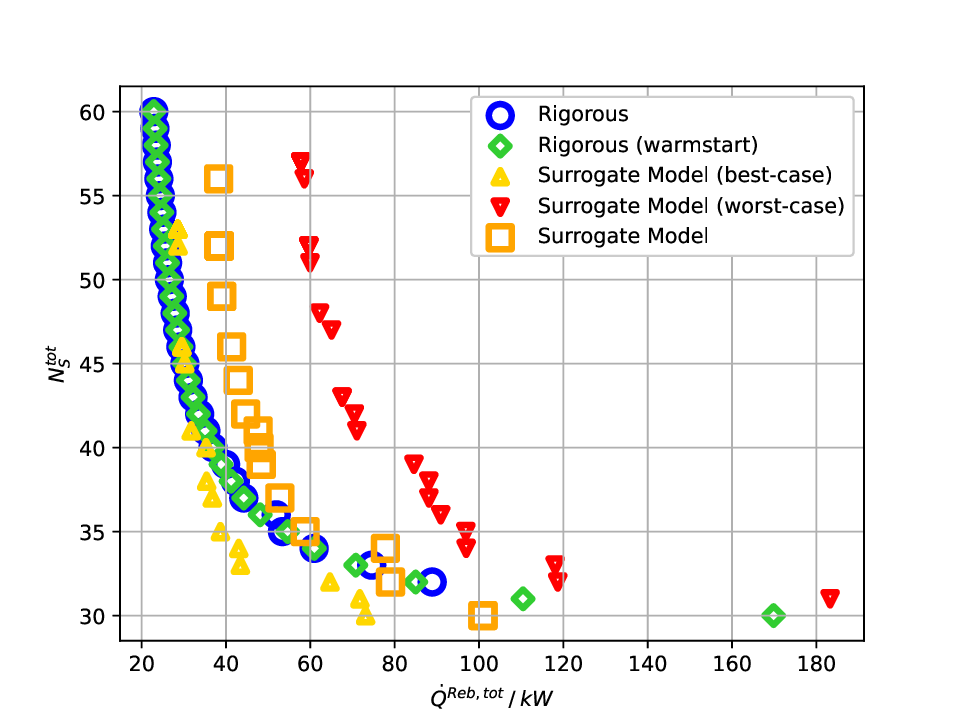}
    \caption{
        NQ curves for Pentanoic acid (CAS 109-52-4) as entrainer in the system Acetone+Pentanoic acid+Chloroform.
        The NQ curves were obtained using the Rigorous (MESH-equation based modeling) for the optimization, using 100 random start locations (blue circles) and the solutions of the surrogate model-based optimizations (green diamonds).
        The NQ curves obtained using the ternary reusable surrogate model for the distillation columns are shown by the orange squares.
        The best- and worst-case results using the surrogate model in the optimizations are shown by the yellow upside-triangles and the red downside-triangles, respectively, while best- and worst-case means adding $\pm 0.5 \cdot \Delta$ of the confidence regions \tabref{tab:confidence_intervals} to the surrogate model predictions.
        % Using surrogate-based solutions as warm-starts improves the discovery of the rigorous Pareto frontier.
    }
    \label{fig:nq_curve_comparison_109-52-4}
\end{figure}

Furthermore, the goal of surrogate modeling usually is not to replace the rigorous modeling.
Instead, the results obtained from surrogate model-based optimization can be used to support the rigorous optimization.
As shown in \figureref{fig:nq_curve_comparison_109-52-4}, using the optimal solutions from the surrogate-based optimization as warm-starts, the rigorous (MESH-equation based) optimization results in an approximation of the NQ curve, including points at a lower total number of column stages than what was found using 100 random multi-starts alone.
Also, in addition to the rigorous NQ curves and the surrogate model-based NQ curve, we computed a surrogate model-based NQ curve representing a worst- and best-case scenario, considering the confidence intervals \tabref{tab:confidence_intervals} of the trained distillation column surrogate model.
Best case means, that we add $0.5 \cdot \Delta \liquidmolarfractionof{\product\column_i}$ and subtract $0.5 \cdot \Delta \reboilerduty_{\column_i}$ of the surrogate model predictions for each column $i$ in the flowsheet during optimization, c.f. (\tabref{tab:confidence_intervals}).
This represents the best possible situation under the calibrated surrogate model confidence regions, where the product purity is overestimated and the reboiler duty is underestimated.
The worst case is computed vice versa.
For the system displayed in \figureref{fig:nq_curve_comparison_109-52-4}, the surrogate model-based NQ curve is too pessimistic and the best case curve is closer to the rigorous NQ curve.
The NQ curves for all entrainer candidates are featured in the Supporting Information to this work.

% However, the surrogate model is not infallible.
% \figureref{fig:nq_curve_comparison_1656-48-0} shows a case, 2-Cyanoethyl ether, where the surrogate's prediction is poor and significantly deviates from the rigorous frontier.
% This highlights an important limitation, which we discuss in the following section.

% \begin{figure}[h!]
%     \centering
%     \includegraphics[width=0.95\textwidth]{figures/nq_curve_comparison_1656-48-0.eps}
%     \caption{
%         NQ curve comparison for 2-Cyanoethyl ether (CAS 1656-48-0) as entrainer in the system Acetone+2-Cyanoethyl ether+Chloroform, showing a significant deviation between the surrogate prediction and the rigorous result.
%     }
%     \label{fig:nq_curve_comparison_1656-48-0}
% \end{figure}

% The complete set of NQ curves for all 37 entrainer candidates is available in the Supporting Information.

\subsection{Discussion: Ranking Accuracy and Model Limitations}
\begin{figure}[h!]
    \centering
    \includegraphics[width=0.7\textwidth]{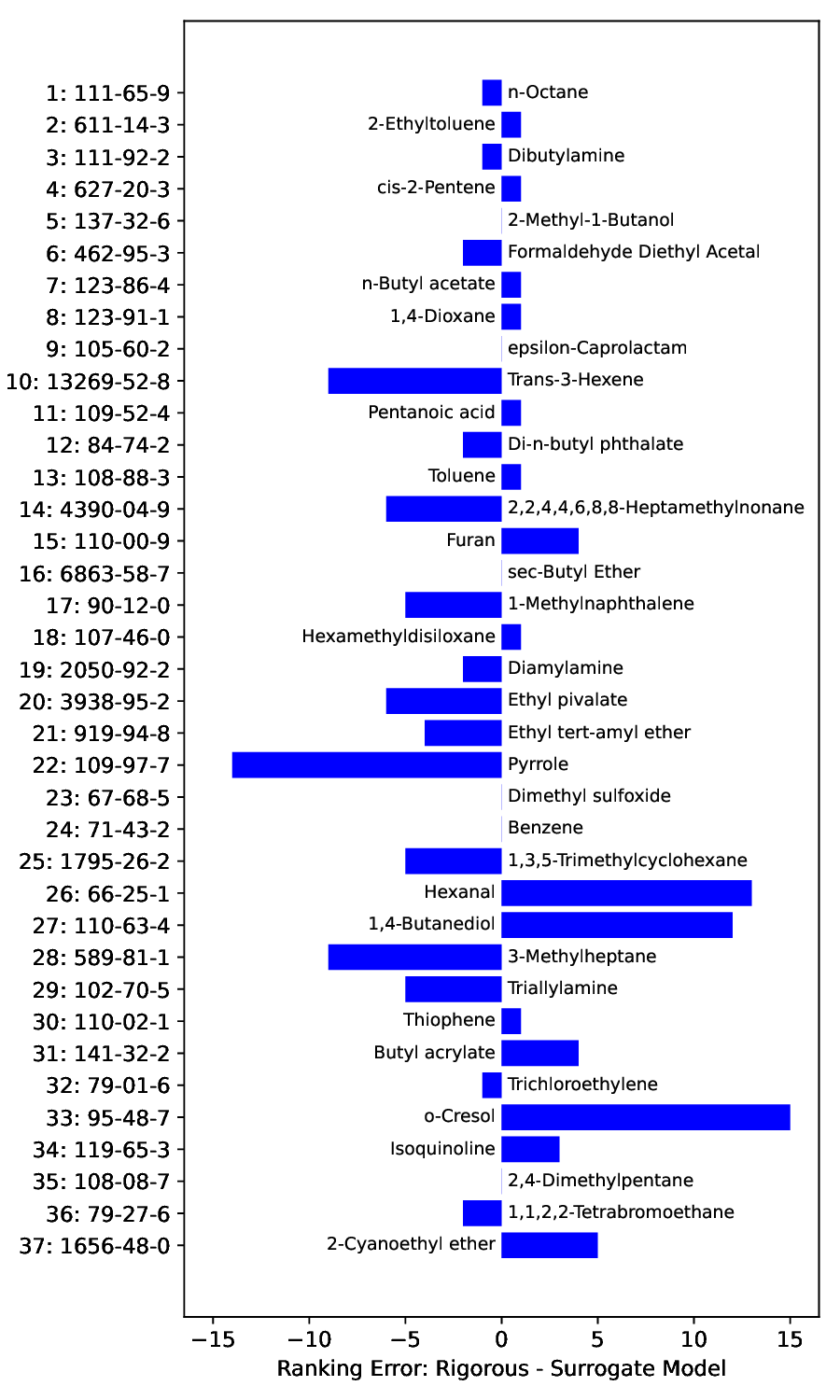}
    \caption{
        Ranking of 37 entrainer candidates based on surrogate-model-generated NQ curves.
        Candidates are ordered from best (top, rank 1) to worst (bottom, rank 37) according to the rigorous optimization-based ranking.
        The horizontal bars indicate the magnitude of the ranking error, defined as the absolute difference between the rank determined from a fully rigorous optimization and the surrogate-based rank.
    }
    \label{fig:nq_curves_pfo_ranking}
\end{figure}

The ultimate goal of the case study is to rank the entrainer candidates based on their predicted performance, thereby identifying the most promising options for further, more rigorous study.
To achieve this, a dominance criterion is established to compare the NQ curves.
An entrainer candidate A is considered superior to candidate B if its Pareto frontier represents a more favorable CAPEX-OPEX trade-off, i.e., achieving a lower total heat duty for a given total number of stages for the majority of points along the frontier.
Using this pairwise comparison, the full set of candidates is sorted to produce the final ranking.
In \figureref{fig:nq_curves_pfo_ranking}, we show the error of that ranking, compared to the results from ranking the NQ curves obtained from fully-rigorous optimization.

The ranking derived from our surrogate model successfully identifies the top-performing entrainers and correctly isolates the low-performing ones.
In particular the most promising candidates (top 10 category in \figureref{fig:nq_curves_pfo_ranking}) are successfully ranked by the surrogate model.
This shows that the trained surrogate model has successfully been used in a reusable manner: The same surrogate model was deployed in a flowsheet consisting of three distillation columns, all operating at different conditions, performing different separations.
The surrogate models were interconnected to form the entire flowsheet and optimization on this interconnected system yields a reasonably close approximation to the true ranking for most of the entrainer candidates.
Still, there are some entrainer candidates that are ranked less precisely than others.
These typically are systems whose modelfluid features lie in a sparse region of the surrogate's training data space.
Also, in the optimization case study used to demonstrate the reusable surrogate model, even small inaccuracies in the predicted product compositions can have a considerable impact on the final NQ curve.
This is because there are regions in the design space where the total reboiler duty is highly sensitive to small changes in process parameters.
% A product composition prediction offset by approximately 3\%, which is within the 95\% confidence interval identified via conformal prediction (\sectionref{sec:modeltraining}), can therefore significantly shift the calculated NQ curve.
% It is worth noting, however, that despite the poor NQ curve prediction, the final ranking error for 2-Cyanoethyl ether was not excessively large (\figureref{fig:nq_curves_pfo_ranking}), suggesting the surrogate still correctly identified it as a low-performer.

The computational effort required for the optimization based on the pre-trained reusable surrogate model is considerably cheaper than that of a full rigorous optimization.
The optimization based on the surrogate model achieves a speedup of approximately two orders of magnitude compared to optimization that requires solving the MESH equations directly.
% While the optimization based on the rigorous (MESH equation-based) distillation column simulation often took several hours to complete (whole NQ curve), the optimization using the surrogate model can be completed in a matter of minutes.

In \appendixref{sec:rankingcomparison}, we compare the ranking accuracy achieved in this work—using optimization based on the reusable surrogate model—with the results reported in \cite{Bubel2025c}.
As discussed in more detail in \appendixref{sec:rankingcomparison}, the authors of \cite{Bubel2025c} employ a different approach, relying on a local approximation of the rigorous entrainer distillation flowsheet derived from the solution of a process fluid optimization problem.

\section{Conclusion and Outlook}
\label{sec:conclusion}
This work addresses a foundational limitation of surrogate modeling in chemical process engineering: the system-specific nature of state-of-the-art models, which restricts their reusability.
We have demonstrated the development of a single, global surrogate model for distillation columns capable of generalizing not only across a wide range of operating and design parameters but, for the first time, across the entire chemical space of homogeneous ternary vapor-liquid mixtures.
This surrogate model can be used in a variety of flowsheets and processes, supporting simulation studies and optimization problems.

This advancement in surrogate model reusability is based on two core contributions.
First, we apply a modelfluid representation in a novel way that enables the surrogate to learn the fundamental relationships between thermodynamic behavior and process performance across the entire mixture space, a scope of generalization not previously achieved for distillation column surrogate models.
Second, and most critically, we employ a powerful ML-for-AI strategy to overcome the data scarcity that has traditionally constrained such ambitious modeling efforts.
This approach leverages an existing machine learning model for property prediction to generate a large and physically representative dataset, which in turn provides the necessary foundation to train our more complex, general-purpose AI-based surrogate.
The outcome is a surrogate model of a distillation column that can be used for any ternary mixture obeying homogeneous vapor-liquid phase behavior.

In a case study on entrainer distillation, we demonstrated the practical benefit of this reusable surrogate.
By providing an explicit mapping from inputs to outputs, the model enables the rapid and robust generation of Pareto-optimal frontiers for an entrainer candidate pool, a screening task that would be computationally expensive with rigorous models alone.
The resulting entrainer ranking showed a strong correlation with rankings derived from rigorous optimization, proving the surrogate's utility as a powerful tool to support and accelerate process synthesis.
Furthermore, the integration of conformal prediction provides statistically rigorous confidence intervals, offering a crucial measure of reliability for engineering applications.
The ability to quantify prediction uncertainty is a key enabler for surrogate models in practical applications.

This work is intended to showcase what is possible for a new generation of surrogate models and to outline a path toward true reusable prediction models for unit operations.
% While the model's accuracy is naturally highest in well-represented regions of the training data, the case study demonstrates its value in exploring vast design spaces.
The goal is not to deliver a final, perfect model, but rather to establish a new methodology and a proof-of-concept for what can be achieved when ML-based data generation is combined with AI-based surrogate modeling.

While this work focuses on ternary mixtures, the modelfluid representation generally extends to multi-component systems.
However, the cost of training data generation and model training increases significantly with the number of components, which poses a challenge for practical implementation.
Also, in order to prevent training separate models for each number of components, an architectural approach to modeling multi-component mixtures would be of high value on the road to broader reusability.
% The success of this work opens clear and promising avenues for future research.
Another crucial next step would be to enhance the ML-for-AI workflow by developing more sophisticated and robust thermodynamic consistency checks for large-scale, synthetically generated fluid systems.
Improving the quality and reliability of the training data at its source is the most direct path to increasing the accuracy and robustness of the final surrogate model.
Furthermore, to manage the risk of the surrogate predicting physically infeasible operating points, future work should explore the development of a complementary feasibility classifier, as demonstrated in \cite{Hoeller2023}.
Integrating such a classifier would ensure that surrogate-based optimizations converge to solutions that are not only mathematically optimal but also physically realizable.

Ultimately, this work presents the potential of pre-trained, globally valid surrogate models that can serve as off-the-shelf digital assets for the process industries.
% By lowering the computational barrier to comprehensive optimization and design space exploration, this new class of reusable surrogate models has the potential to significantly accelerate innovation and improve the efficiency and robustness of chemical process development.

\appendix
\renewcommand{\theequation}{\thesection.\arabic{equation}}

\section{VLE Consistency Checks} \label{sec:vlechecks}
This section details the set of rules used to classify the vapor-liquid equilibrium (VLE) behavior of synthetically generated binary mixtures as thermodynamically consistent or inconsistent.
These checks are a critical step in our data generation workflow, ensuring the quality and physical realism of the dataset used to train the reusable surrogate model.

The necessity for these checks arises from the empirical nature of the thermodynamic property models commonly used in process simulation, including those in this work.
Models for vapor pressure (e.g., Antoine equation) and liquid-phase activity coefficients (e.g., Margules model) are not derived from first principles and can be parameterized to describe behavior that is not physically attainable.
When these models are parameterized using ML-predicted data, as done in our large-scale dataset generation (\sectionref{sec:dataset}), the risk of producing such thermodynamic artifacts increases.
Therefore, a robust filtering mechanism is required to discard these unphysical systems.

Our workflow involves generating a vast number of potential ternary systems by combining pure component data from the DIPPR 801 database \cite{Wilding1998} with ML-predicted activity coefficients at infinite dilution \cite{Jirasek2020, Damay2021}.
Before a ternary system is accepted, we analyze the VLE of its three constituent binary sub-systems.
If any of the binary pairs fail one or more of the checks described below, the entire ternary system is removed from the dataset.
While the checks are discussed for isobaric conditions, many of their principles apply to isothermal systems as well.

The set of checks presented in this section reflects the current state of our research and provides a robust filter for the large-scale dataset generation.
It is important to note, however, that this collection of rules is based on a combination of thermodynamic principles and empirical heuristics, and does not constitute a formal proof of thermodynamic consistency.
Consequently, it is possible that some systems passing these checks may still exhibit VLE profiles that appear atypical or thermodynamically questionable, even if they do not violate a specific criterion defined below.

Figures illustrating systems that violate these checks are provided in the Supporting Information to this work.

\subsection{Endpoint Convergence Check}
This check ensures that the VLE curves correctly converge to the pure component endpoints.
It is a fundamental, physically motivated test to prevent artifacts at the composition boundaries.
The conditions checked at the limits of the mole fraction $\liquidmolarfractionof{i}$ are:
\begin{itemize}
    \item $\vapormolarfractionof{i}\left(\liquidmolarfractionof{i} \to 0\right) = 0$
    \item $\vapormolarfractionof{i}\left(\liquidmolarfractionof{i} \to 1\right) = 1$
\end{itemize}
Given the nature of the isobaric VLE calculation, the bubble and dew temperatures are guaranteed to match the pure component boiling points at these limits.

\subsection{Continuity Check}
This check is motivated by the empirical observation that some parameterizations lead to unphysical discontinuities, or jumps, in the VLE curve.
It identifies systems where an infinitesimally small change in liquid composition causes a large, finite change in the corresponding vapor composition.
Since such behavior is not observed in real fluid systems, we implement this check with a configurable tolerance to filter out these artifacts.
This is a purely heuristic check and the threshold needs to be set by the user.
In this work, we use a threshold of $\vert\vapormolarfractionof{i}^{(j+1)} - \vapormolarfractionof{i}^{(j)}\vert \leq 0.2 \molefracunit$, where superscripts $(j+1)$ and $(j)$ denote two consecutive points on the VLE curve.

\subsection{Monotonicity Check}
This check enforces that the bubble and dew curves are monotonic in regions where physical principles require them to be.
For a simple binary system, the vapor mole fraction $\vapormolarfractionof{i}$ must be a monotonically increasing function of the liquid mole fraction $\liquidmolarfractionof{i}$.
Violations often appear as non-physical hooks or reversals in the VLE diagram and frequently co-occur with other artifacts.

\subsection{Phase Envelope Check}
This check enforces a direct consequence of the second law of thermodynamics: at any given composition, the dew point temperature must be greater than or equal (singular VLE points) to the bubble point temperature.
A direct comparison can be complicated because the standard VLE calculation yields bubble temperatures at discrete liquid compositions and dew temperatures at discrete vapor compositions, while the liquid and vapor compositions usually differ.
To perform the check, we interpolate the bubble and dew temperatures onto a common composition axis and ensure that $T^{\text{dew}} \ge T^{\text{bubble}}$ across the entire range.
A small tolerance is permitted to account for numerical noise, especially near azeotropes.
In this work, we tolerate violations of this criterion in the range of $\pm 0.05 \molefracunit$ of an azeotropic point.

In fact, there exists the possibility to create VLEs satisfying extended Raoult's Law where for all concentrations visited, the bubble temperature is larger than the dew temperature.
Using the modelfluid features of a binary system and replacing the saturated vapor temperatures of the pure components generates such a system.

\subsection{Azeotrope and Extrema Consistency Check}
This check combines several criteria to ensure the geometric shape of the VLE diagram is physically consistent, particularly for systems with one or more azeotropes.
It validates the relationship between azeotropes and the extrema of the temperature curves.
The check proceeds as follows:
\begin{itemize}
    \item \textbf{Azeotrope Counting:} The number of azeotropes is determined by counting the sign changes of $(\vapormolarfractionof{i}\inb{\liquidmolarfractionof{i}} - \liquidmolarfractionof{i})$ of a series solutions of extended Raoult's Law for equidistantly monotonically increasing values of $\liquidmolarfractionof{i}$.
    \item \textbf{Extrema Matching:} The number of extrema (maxima or minima) in the bubble and dew temperature curves is counted and must match the number of azeotropes.
    \item \textbf{Polyazeotropy Nature:} For systems with multiple azeotropes, it enforces rules on their sequence to filter out unphysical combinations (e.g., see \cite{Anjum2024}).
\end{itemize}
A violation of any of these conditions indicates an inconsistent VLE shape.

\subsection{Zeotropic Curvature Check}
This check is designed specifically for zeotropic systems to detect unphysical S-shapes in the temperature-composition curves.
Such shapes, characterized by an inflection point in the central composition range, are often indicators of (incipient) liquid-liquid immiscibility, which is not captured by the VLE model being used.
The check computes the second derivative of the bubble and dew curves and flags any system with a sign change (inflection point) in the core region.

In this work, we use a tolerance of $0.1 \molefracunit$ of the pure component locations as many systems seemed to fail this check near the pure component endpoints, while generally exhibiting a smooth and consistently looking VLE.

\subsection{Numerical Validity Check}
This final, straightforward check ensures data integrity by scanning all computed VLE data points for non-numerical values, such as \texttt{NaN} or \texttt{Inf}.
The presence of such values indicates a failure in the numerical VLE solver and results in the system being discarded.

\section{Comparison of ranking accuracy to related work}
\label{sec:rankingcomparison}
\begin{figure}[h!]
    \centering
    \includegraphics[width=0.7\textwidth]{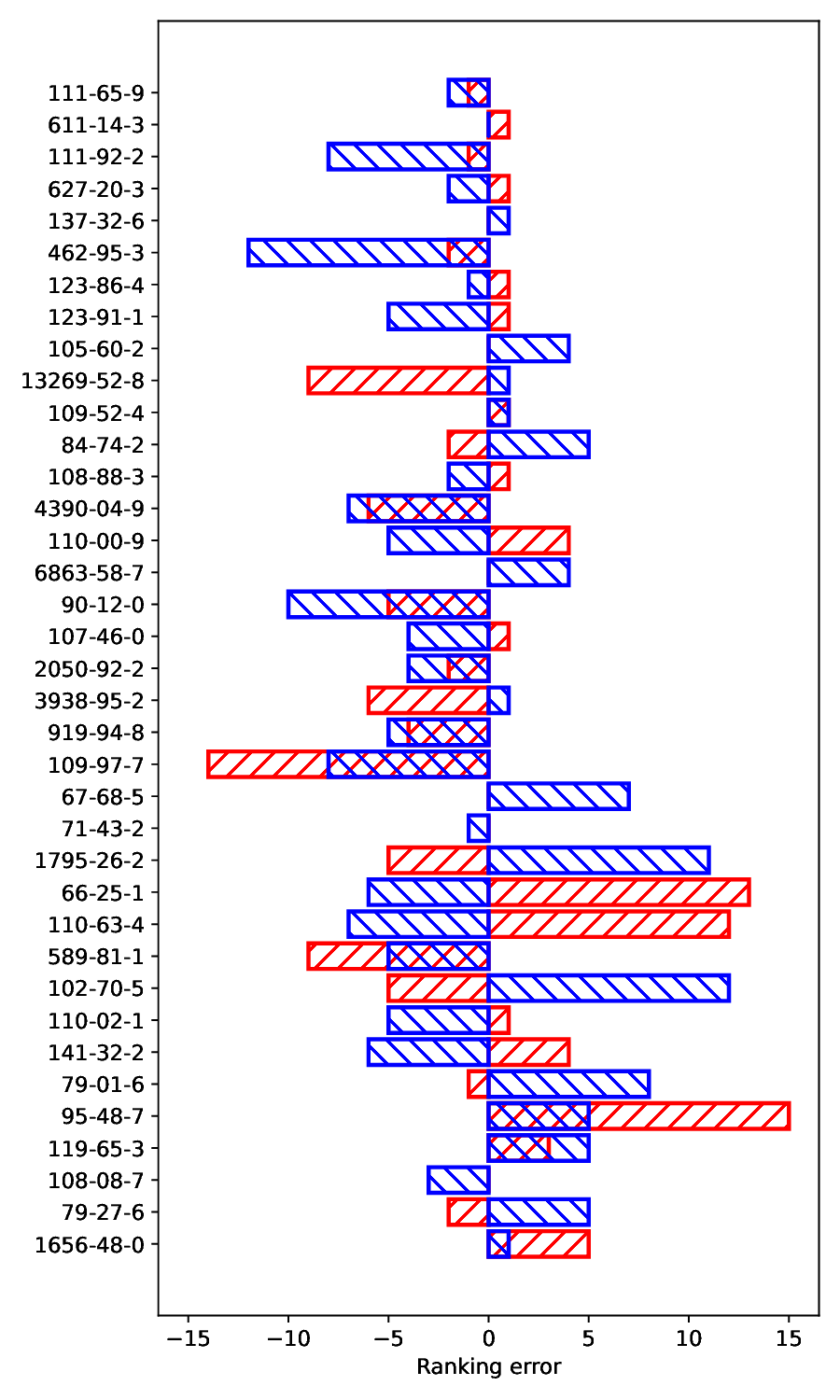}
    \caption{
        Comparison of the error using surrogate model-based ranking of NQ curves from this work to the error of the approximate ranking presented from \cite{Bubel2025c}.
        Candidates are ordered from best (top, rank 1) to worst (bottom, rank 37) according to the rigorous optimization-based ranking.
        Red bars ("//") show the ranking error of this work (surrogate modeling); blue bars ("\textbackslash{}\textbackslash{}") show the ranking error of \cite{Bubel2025c}.
        The horizontal bars indicate the magnitude of the ranking error, defined as the absolute difference between the true rank obtained from the rigorous NQ curves and the approximate ranking obtained using the surrogate model-based ranking or the one from \cite{Bubel2025c}.
    }
    \label{fig:nq_curves_pfo_ranking_pfo_comparison}
\end{figure}

In addition to the discussion of the ranking error using surrogate model-based optimization in \sectionref{sec:ranking}, we compare the entrainer ranking results achieved in this work, by the use of the reusable surrogate model, to those presented in \cite{Bubel2025c}, who use the same case study, in \figureref{fig:nq_curves_pfo_ranking_pfo_comparison}.

In \cite{Bubel2025c}, the authors used a first-order approximation of the objective function ($\totalreboilerduty$), at the solution to their process fluid optimization problem.
In that work, they did not only optimize the process specifications of the columns in the flowsheet but also the modelfluid features that describe the entrainer component or its interactions with the solutes.
The results achieved in this work are not only more accurate but come with a significantly lower computational effort compared to the rigorous process fluid optimization conducted in \cite{Bubel2025c}.
In \figureref{fig:nq_curves_pfo_ranking_pfo_comparison}, we show the differences (error) in the entrainer candidate ranking from the rigorous approximation used in \cite{Bubel2025c} and the surrogate model-based ranking presented in this work.
It is found that the majority of systems are ranked in close vicinity while the surrogate model-based ranking outperforms the results from \cite{Bubel2025c} in identifying the best-performing entrainers.
For the mid- and low-performing candidates, the ranking error varies between the two approaches.
The mean value over the absolute ranking error for all entrainer candidates is $3.73$ for the surrogate model-based ranking and $4.84$ for the ranking from \cite{Bubel2025c}.

A further important consideration arises from the differing domains of the rigorous and surrogate models.
Rigorous simulators can fail to converge for certain input specifications, effectively creating holes or infeasible regions in the design space.
The surrogate model, by its nature as an explicit function, will always return a prediction for any given input.
This is favorable for an optimization algorithm, avoiding convergence failures that can prematurely terminate rigorous optimization runs.
However, this robustness also introduces a critical challenge.
The optimizer may identify an optimal point in a region where the rigorous model would not have a feasible solution, as the surrogate is trained only on data from successful simulations and has no inherent knowledge of this feasibility boundary.
While this behavior was not observed to be a dominant issue in this work, it remains a crucial aspect to manage when applying surrogate models in practice.
A robust strategy to mitigate this risk is to complement the regression surrogate with a classification model trained to predict process feasibility, an approach that is discussed further in the outlook.
Also, the optimization problem using the surrogate model may be different to that based on the rigorous modeling.
An immediate effect of this may be that convex NQ curves using the rigorous modeling can become non-convex using the surrogate models.
This can be observed by the NQ curves for some of the entrainer candidates presented in the Supporting Information to this work.

\section{Distance Metrics for Pareto Frontier Comparison} \label{sec:distance_metrices}
In this section we describe in distance measures used to compare the Pareto frontiers obtained from the surrogate model-based optimizations to the reference frontier using the rigorous modeling of the distillation columns in the flowsheet.

We use the generational distance (GD) and inverted generational distance (IGD) measures, as described in \cite{Ishibuchi2015}, to assess the proximity of the surrogate model-based frontiers to that of the rigorous optimization and vice versa.
We also use the average symmetric distance (ASD), which is the mean value of GD and IGD, and the Hausdorff distance to provide a more comprehensive comparison.
For all of these distance measures, a lower value indicates a better approximation of the true Pareto frontier.
The GD metric \eqref{eq:gd}, which measures the average distance of the Pareto points of the surrogate model-based optimization to their respective closest-neighbor on the rigorous Pareto frontier, may not reveal if the surrogate model-based frontier does not cover the entire frontier from the rigorous optimization.
On the other hand, the IGD metric \eqref{eq:igd}, which measures the average distance of the Pareto points of the rigorous optimization to their respective closest-neighbor on the surrogate model-based frontier, may not reveal if the surrogate model-based frontier contains any points that are far away from the rigorous Pareto frontier.
This is why we also use the average symmetric distance (ASD) \eqref{eq:asd}, which yields the average of both the GD and IGD metrics.
In addition, we use the Hausdorff distance metric \eqref{eq:hausdorff}, which measures the maximum distance of a point on one frontier to its closest-neighbor on the other frontier.
This metric can be used as a warning flag, indicating when there are outliers on any of the curves in the comparison, both for the surrogate model-based and the rigorous optimization.

Since, we are only interested in the geometric distance between the approximate frontiers and the reference (rigorous) frontiers, we refrain from using any metrics that consider dominance of frontiers, such as hypervolume or IGD+ \cite{Ishibuchi2015}.

For the distance measures described in the following, we let \newline
$P_{\text{true}} = \{t_1, t_2, ..., t_{|P_{\text{true}}|}\}$ be the set of points representing the true Pareto front, and let $P_{\text{approx}} = \{a_1, a_2, ..., a_{|P_{\text{approx}}|}\}$ be the set of points representing an approximate front.
When using the distances in \sectionref{sec:applications}, we consider the NQ curves obtained from the optimization on the rigorous distillation column model as the true front, and that NQ curves from the surrogate model-based optimizations as the approximate front.
The distance between any two points $u$ and $v$ in the objective space is given by the Euclidean distance $d(u,v)$.
The below formulas and definitions are obtained from \cite{Ishibuchi2015} and the references therein.

\paragraph{Generational Distance (GD)}
The GD metric measures the average distance from each point in the approximate front to its nearest neighbor in the true front.
It quantifies how close the found solutions are to the true solutions. A lower value is better.
\begin{equation} \label{eq:gd}
    \text{GD}(P_{\text{approx}}, P_{\text{true}}) = \frac{1}{|P_{\text{approx}}|} \sum_{a \in P_{\text{approx}}} \min_{t \in P_{\text{true}}} d(a, t)
\end{equation}

\paragraph{Inverted Generational Distance (IGD)}
The IGD metric measures the average distance from each point in the true front to its nearest neighbor in the approximate front.
It is a comprehensive metric that evaluates both the convergence and the diversity (coverage) of the approximate front. A lower value is better.
\begin{equation} \label{eq:igd}
    \text{IGD}(P_{\text{approx}}, P_{\text{true}}) = \frac{1}{|P_{\text{true}}|} \sum_{t \in P_{\text{true}}} \min_{a \in P_{\text{approx}}} d(t, a)
\end{equation}

\paragraph{Average Symmetric Distance (ASD)}
The ASD, provides a balanced, symmetric measure of the average geometric distance between the two fronts.
It is simply the mean of the GD and IGD metrics.
A lower value indicates a better overall geometric match.
\begin{equation} \label{eq:asd}
    \text{ASD}(P_{\text{approx}}, P_{\text{true}}) = \frac{\text{GD}(P_{\text{approx}}, P_{\text{true}}) + \text{IGD}(P_{\text{approx}}, P_{\text{true}})}{2}
\end{equation}

\paragraph{Hausdorff Distance}
The Hausdorff distance ($d_H$) measures the maximum worst-case distance between the two fronts.
It is defined as the maximum of two directed distances, where each directed distance finds the point in one set that is farthest from any point in the other set.
It is highly sensitive to outliers. A lower value is better.
\begin{align} \label{eq:hausdorff}
    d_H&(P_{\text{approx}}, P_{\text{true}})=\\ \nonumber
    &\max \left( \max_{a \in P_{\text{approx}}} \left\{ \min_{t \in P_{\text{true}}} d(a, t) \right\}, \max_{t \in P_{\text{true}}} \left\{ \min_{a \in P_{\text{approx}}} d(t, a) \right\} \right)
\end{align}

% \bibliographystyle{elsarticle-num}
% \bibliography{literature}

\end{document}